\documentclass[journal]{IEEEtran}
\usepackage{amsmath,amssymb,bm}
\usepackage{graphicx}
\usepackage{booktabs}
\usepackage{cite}
\usepackage[per-mode=symbol]{siunitx}
\usepackage{url}
\usepackage{enumitem}
\usepackage{tikz}
\usetikzlibrary{arrows.meta,positioning,calc,fit}
\newtheorem{proposition}{Proposition}
\newcommand{\shat}{\hat{\bm{s}}}
\newcommand{\rv}{\bm{r}}
\newcommand{\av}{\bm{a}}
\newcommand{\bv}{\bm{b}}
\newcommand{\cv}{\bm{c}}
\newcommand{\Mm}{\bm{M}}
\newcommand{\Cm}{\bm{C}}
\newcommand{\RQ}{\bm{R}_{Q}}
\newcommand{\Id}{\bm{I}}
\newcommand{\Gt}{\overline{\overline{\bm{G}}}}
\newcommand{\sm}{\hat{\sigma}}
\newcommand{\Tr}{\operatorname{Tr}}
\newcommand{\pv}{\hat{\bm{p}}}

\begin{document}


\title{A Theory of Transmit-Receive-Manifold of Single-Photon Quantum-Dot Antenna Arrays}

\author{Angeru~Pavan~Kumar,~\IEEEmembership{Student~Member,~IEEE,}
        and~Sandip~Ghosal,~\IEEEmembership{Member,~IEEE}%
\thanks{\color{red}This work has been submitted to the IEEE for possible publication.
Copyright may be transferred without notice, after which this version may no
longer be accessible.}

}

\markboth{IEEE Journal Latex}%
{A. P. Kumar and S. Ghosal: A Theory of Transmit-Receive-Manifold of Single-Photon Quantum-Dot Antenna Arrays}

\maketitle
\raggedbottom

\begin{abstract}
A quantum-dot array that receives a single photon is a collectively coupled antenna, and its receive manifold is not always the plane-wave manifold. So, this work aims to develop a complete transmit-channel-receive manifold model for quantum dot (QD) arrays. For the sake of validation, it considers a reference target pattern which can be obtained using classical array excitation. However, the present work proposes single photon excitation model to a 
$4\times4$ InAs/InP arrays emitting at 1550 nm to generate the same target pattern. In the receiving end, it considers the same 16-element QD array configuration, identical to the transmitter side. The channel between the transmitter and receiver is modeled through the dyadic Green's tensor formulation. With the single photon excitation, the C-band lifetime and dephasing parameters are compared against the experimentally measured results from the literature. In spite difference in physical excitation, the steered beam of the first-order QD array model appears equivalent to a classical phased array. The new result concerns reception: a lifetime-limited photon is not narrowband relative to the collective receiver modes, so, after the temporal (equivalently spectral) degree of freedom is traced out, it leaves the dipole-coupled array in a mixed spatial state of purity 0.949, and the standard narrowband coupling-aware manifold then errs by 1.81 degrees, worse than applying no correction. A spectrally matched (SM) manifold based on the photon spectrum reduces the model-mismatch error to 0.031 degrees, while finite-snapshot Monte Carlo results approach the corresponding exact and Gaussian Cramer-Rao bounds. These results demonstrate the importance of incorporating finite photon bandwidth and collective receiver coupling when modeling single-photon reception and estimating arrival direction.The proposed mutual-couping aware quantum formalisms will be useful in next generation communication of Telecom C-band applications.

\end{abstract}


\section{Introduction}\label{sec:intro}
\begin{figure*}[!t]
\centering
\includegraphics[width=1\textwidth]{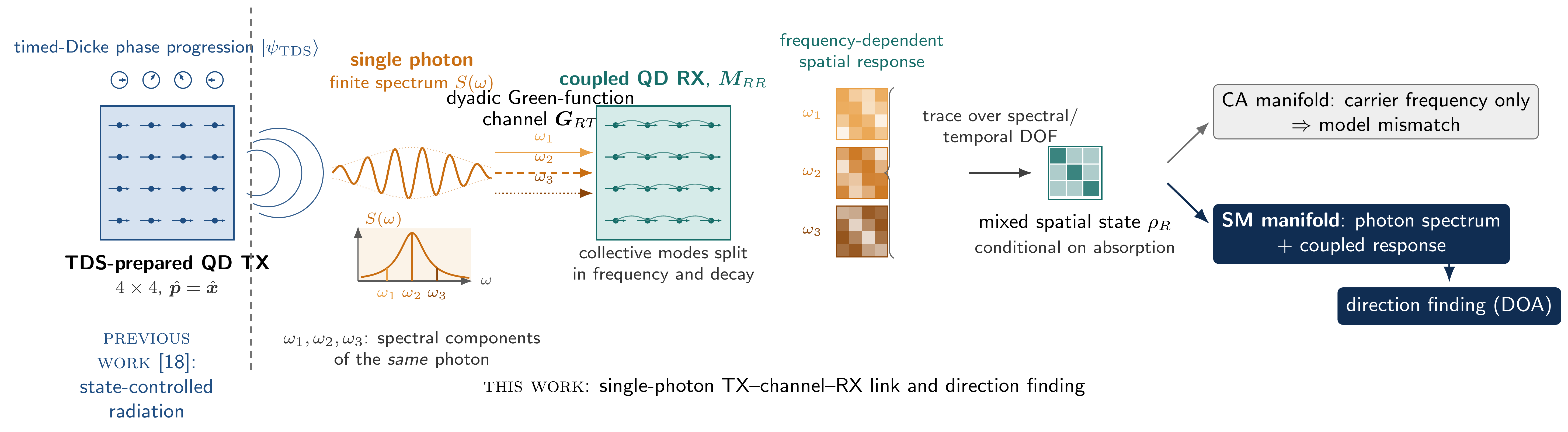}
\caption{Conceptual framework of the quantum-antenna link. Previous work \cite{prevwork} established state-controlled directional radiation from a timed-Dicke-state QD array. The present work extends this problem to a single-photon TX--channel--RX link, where the photon spectrum $S(\omega)$ is set by the transmitter lifetime and propagated through the dyadic Green-function channel $\bm{G}_{RT}$. The collectively coupled receiver $\Mm_{RR}$ produces frequency-dependent spatial responses, here illustrated for three spectral components $\omega_{1},\omega_{2},\omega_{3}$ of the same photon; tracing over the spectral/temporal degree of freedom yields the conditional receiver spatial state $\rho_{R}$. This motivates a spectrally matched (SM) manifold rather than the conventional carrier-frequency coupling-aware (CA) manifold for direction finding.}
\label{fig:concept}
\end{figure*}

\IEEEPARstart{I}{n} a conventional array, the pattern is controlled by the feed currents: an external network imposes a complex excitation weight $w_{n}$ on the $n^{th}$ element. In an array of quantum emitters, the pattern is rather dependent on the state of the emitters, because $w_{n}$ is replaced by the lowering operator $\sm_{n}^{-}$ of the $n^{th}$ emitter and the density matrix becomes the reconfiguration variable \cite{mokhlespour2012,slepyan2013,mikhalychev2018,slepyan2020}. The radiation mechanism is collective spontaneous emission \cite{dicke1954,lehmberg1970,ficek2002}. One collective state is directly useful as an antenna excitation: the timed Dicke state (TDS), in which one excitation is shared by all emitters with a progressive phase so that the emission is directed along a chosen wave vector \cite{scully2006,svidzinsky2010}. At sub-wavelength spacing, the dipole--dipole coupling between emitters becomes strong and geometry dependent, splitting the collective modes in both frequency and decay rate \cite{asenjo2017,masson2022,shahmoon2017,rui2020}. In semiconductor quantum dots (QDs), this regime has been reached only after the dots were tuned into mutual resonance \cite{kim2018,grim2019,tiranov2023,abdussalam2012}.\\
In our previous work \cite{prevwork}, we proposed a theoretical framework for the state-controlled directional emission from one- and two-dimensional QD antenna arrays. The earlier work of \cite{prevwork} related the timed Dicke state to the radiated pattern, retained dipole--dipole coupling in the collective decay, separated directional pattern asymmetry from Lorentz reciprocity, and derived a quantum spatial-coherence description with a Fisher-information analysis of direction finding. Two of its conclusions carry into the present paper: the first-order pattern of any pure single-excitation state is reproducible by a classical phased array with the corresponding complex weights, so the non-classical content of such an array resides in its photon statistics rather than in its first-order pattern. That work, however, treated the QD array as a radiating system. In \cite{prevwork}, it did not address what spatial state would be produced when the emitted single photon is received by a second, mutually coupled QD array. The present paper addresses the receive-side problem. The step is therefore not an increase in array dimensionality; it is the transition from state-controlled radiation to a complete quantum-antenna link comprising transmission, propagation, reception, and direction finding.\\
The key problem statement this paper aims to address can be summarized as:\\
\textit{If a TDS-prepared QD array emits a directional single photon and that photon reaches a second dipole-coupled QD array, what spatial state does the receiver acquire, and which receive manifold should its direction finder use?} The intuitive configuration is shown in Fig.~\ref{fig:concept}.

The telecom C-band makes the question concrete. InAs/InP QDs emitting directly at \SI{1550}{\nano\meter} now reach coherence times above \SI{1}{\nano\second} \cite{anderson2020} and resonance fluorescence with near-Fourier-limited coherence \cite{wells2023}, following the first single-photon demonstrations at this wavelength \cite{benyoucef2013,muller2018}; C-band sources have also been used over deployed fibre \cite{yang2024,morrison2023}, and reviews are available in \cite{yu2023,vajner2022}. Those measurements, rather than an idealized confinement model, fix the radiative lifetime and dephasing rates used throughout this paper.\\
Fig.~\ref{fig:concept} shows the physical transmit-receive manifold. A transmitter (TX) array prepared in a timed Dicke state radiates a directional single photon. The photon crosses a free-space channel described by the dyadic Green tensor between the transmitting and receiving dots. A second QD array can absorb that photon; because its dots are mutually dipole-coupled, its response is not that of an uncoupled plane-wave aperture. Conditioned on absorption, the receiver is described by a spatial state $\rho_{R}$, which is read out and used to estimate the direction of arrival. The photon carries two degrees of freedom, and the interaction between them is the core of the problem. Its spatial degrees of freedom are determined by the state-controlled radiation and the propagation direction. Its spectral degree of freedom is set by the finite radiative lifetime $T_{1}$ of the emitting dot, which gives a linewidth of order $\Gamma=1/T_{1}$.\\
The coupled receiver responds differently at different frequencies, so the spectral components of a single photon drive distinct spatial response vectors. When the detector does not resolve the arrival time, that degree of freedom is traced out, and what remains is a spatial density matrix that is, in general, mixed. So, a finite lifetime does not destroy spatial coherence; rather, it produces a finite bandwidth, a frequency-selective receiver maps each spectral component to a different spatial response, and tracing over the spectrum therefore leaves a mixed spatial state.\\
For a classical narrowband array, mutual coupling is handled by building the manifold from the coupling matrix at the carrier \cite{friedlander1991}. In the notation used throughout this paper, the free-space array vector $\av(\shat)$ has elements $[\av(\shat)]_{n}=e^{-ik_{0}\shat\cdot\rv_{n}}$, with $\shat$ the unit direction, $\rv_{n}$ the position of emitter $n$ and $k_{0}=2\pi/\lambda_{0}$; the receiver (RX) collective coupling matrix $\Mm_{RR}$ collects the dot detunings, the coherent dipole--dipole shifts, and the dissipative collective decay. The matrix $\Mm_{RR}$ is itself frequency independent. The frequency dependency nature arises through the resolvent, giving the coupled response
\begin{equation}
\bv(\shat,\omega)=(\omega\Id-\Mm_{RR})^{-1}\av(\shat),
\label{eq:intro_resolvent}
\end{equation}
of which the conventional narrowband coupling-aware (CA) manifold is the single-frequency evaluation
\begin{equation}
\bv_{\mathrm{CA}}(\shat)=(\omega_{c}\Id-\Mm_{RR})^{-1}\av(\shat).
\label{eq:intro_ca}
\end{equation}
The incident free-space phase distribution is described by $\av$, while the inverse receiver operator maps that distribution into the actual coupled-array response. Equation \eqref{eq:intro_ca} is adequate only while the excitation occupies a band narrow compared with the splitting of the collective modes, which a lifetime-limited photon does not. Wideband array processing addressed frequency-dependent coupling before, through focusing-matrix and coherent-subspace methods \cite{wang1985} and through explicitly frequency-dependent coupling matrices \cite{li2017}. In those formulations, the signal spectrum and the coupling matrix were considered as classical array-processing inputs. The present formulation instead derives the receive spatial state from a single-photon TX--Green-function--RX model and retains the spectral--spatial correlations of a single photon until the spectral degree of freedom is traced out.\\
Based on the preceding discussion, three modeling gaps are identified as:
\begin{enumerate}[leftmargin=*,topsep=1pt,itemsep=0pt,parsep=1pt]
\item \emph{Quantum-excitation gap:} The receiver is excited by a single photon, whose mean field vanishes \cite{glauber1963}, rather than by a classical coherent field, so the validity of a coupling-aware manifold must be established rather than assumed.
\item \emph{Electrodynamic-consistency gap:} The element pattern and the dipole--dipole coupling must come from the same electromagnetic model, or the superposition rule linking the radiated pattern to the collective decay matrix is violated.
\item \emph{Finite-bandwidth gap:} A carrier-frequency manifold cannot, in general, represent a lifetime-limited photon interacting with frequency-split collective receiver modes.
\end{enumerate}

\begin{table}[ht!]
\caption{Comparison of This Work With Closely Related Studies}
\label{tab:gap}
\centering
\scriptsize
\setlength{\tabcolsep}{1.6pt}
\begin{tabular}{@{}lcccccc@{}}
\toprule
Work & C-band & Meas. & Coupled & 1-photon & DOA & CRB\\
 & & param. & TX/RX & RX & & \\
\midrule
Si optical phased arr. \cite{sun2013,poulton2017} & \checkmark & -- & -- & -- & -- & --\\
Quantum nanoantennas \cite{mokhlespour2012,slepyan2013} & -- & -- & TX & -- & -- & --\\
Correlation antennas \cite{mikhalychev2018,slepyan2020} & -- & -- & TX & -- & -- & --\\
Atom arrays \cite{asenjo2017,shahmoon2017,rui2020} & -- & \checkmark & TX/RX & -- & -- & --\\
QD superradiance \cite{grim2019,tiranov2023} & -- & \checkmark & TX & -- & -- & --\\
C-band QD sources \cite{wells2023,yang2024} & \checkmark & \checkmark & -- & -- & -- & --\\
DOA with coupling \cite{friedlander1991} & -- & -- & RX & -- & \checkmark & --\\
Wideband DOA + coupling \cite{wang1985,li2017} & -- & -- & RX & -- & \checkmark & \checkmark\\
\textbf{This work} & \checkmark & \checkmark & TX/RX & \checkmark & \checkmark & \checkmark\\
\bottomrule
\end{tabular}
\end{table}

\begin{figure*}[!t]
\centering
\includegraphics[width=0.75\textwidth]{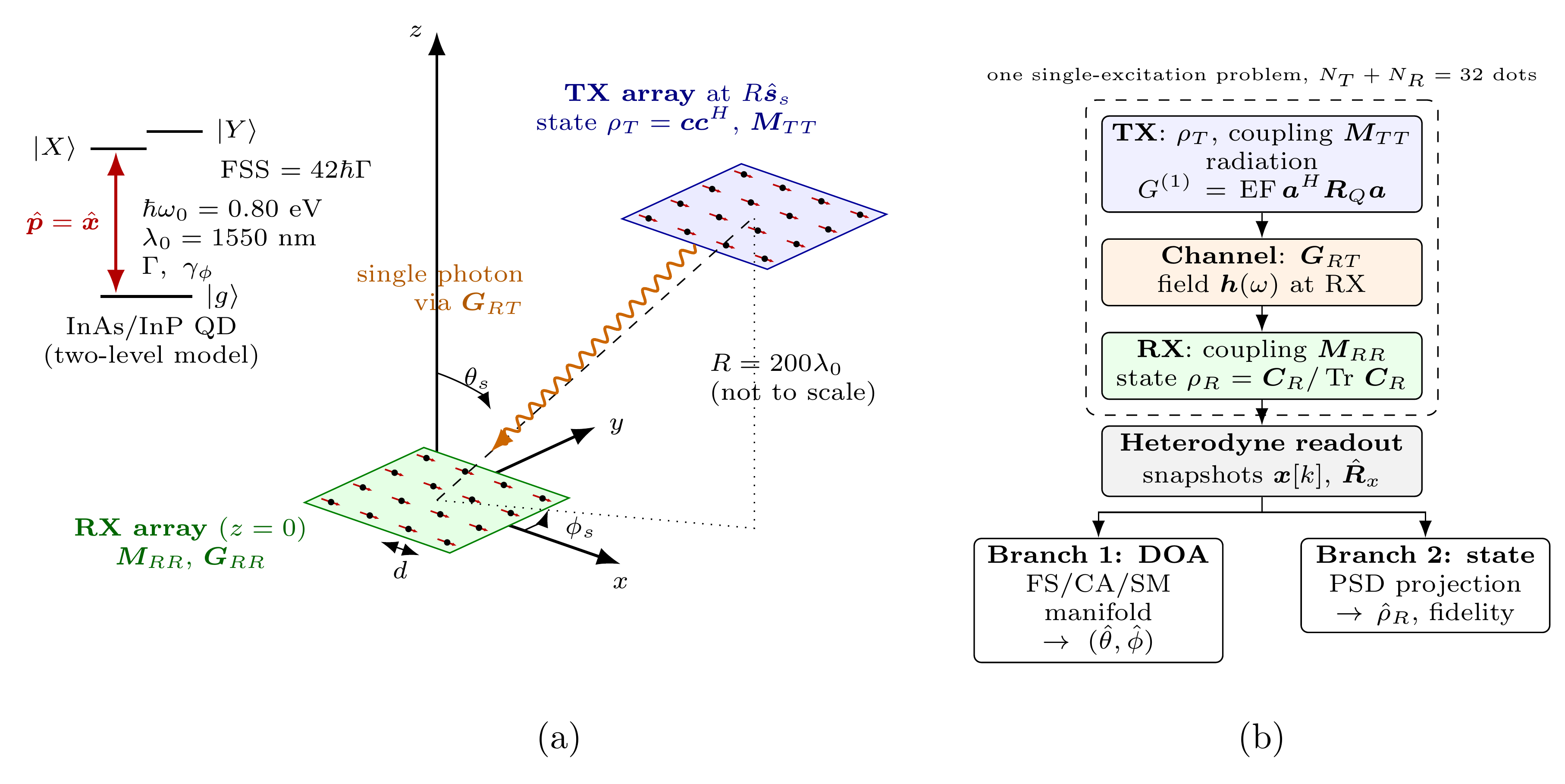}
\caption{System model. (a) Link geometry: the $4\times4$ RX array lies in $z=0$ and the parallel TX array is centred at $R\shat_{s}$, with $\shat_{s}$ given by $(\theta_{s},\phi_{s})$; every dot is an in-plane $\hat{\bm{x}}$ dipole (red arrows) and $d=\lambda_{0}/2$. Inset: two-level model of the dot, with the $\hat{\bm{x}}$-polarized exciton $|X\rangle$ addressed and $|Y\rangle$ lying $42\hbar\Gamma$ away. (b) Signal flow $\rho_{T}\rightarrow\bm{G}_{RT}\rightarrow\rho_{R}\rightarrow\bm{x}[k]$: TX, channel and RX are solved as one collectively coupled single-excitation problem, and the heterodyne snapshots feed two separate inference branches, direction finding and state reconstruction. FS, CA and SM are the manifolds of Section~\ref{sec:doa}.}
\label{fig:arch}
\end{figure*}

This paper addresses these research gaps within a single self-consistent single-excitation model that includes state-controlled transmitter radiation, dyadic Green-function propagation, a collectively coupled receiver, the finite photon bandwidth, the resulting receiver spatial state, and the spectrally matched (SM) receive manifold, through to direction finding and its Cram\'er--Rao bounds. Table~\ref{tab:gap} places the work beside the closest earlier studies, which treat emitter arrays as transmitters \cite{mokhlespour2012,slepyan2013,mikhalychev2018,slepyan2020}, characterize collective modes without an antenna manifold \cite{asenjo2017,shahmoon2017,rui2020,grim2019,tiranov2023}, or assume a classical narrowband field \cite{schmidt1986,capon1969,roy1989,friedlander1991,vantrees2002}. The first-order pattern is classically equivalent, and a classical receiver supplied with the same frequency-dependent manifold would estimate directions equally well. The novelties of contribution are the physically consistent receive-side treatment and the resulting manifold. The arrays and the link constitute a theoretical architecture for numerical investigation, while only the QD parameters are measured values taken from the existing literature.\\
The rest of the paper is organized as follows. Section~\ref{sec:model} describes the architecture and the C-band QD model, and Section~\ref{sec:tx} treats the radiation of the transmitter. Sections~\ref{sec:channel} and \ref{sec:rx} develop the Green's function-based channel and the coupled receiver, and Section~\ref{sec:meas} gives the measurement model. Section~\ref{sec:doa} presents the direction-finding methods, Sections~\ref{sec:results} and \ref{sec:robust} report the numerical results and the robustness study, Sections~\ref{sec:cq} and \ref{sec:disc} discuss the classical--quantum comparison and the limitations, and Section~\ref{sec:concl} concludes the paper.

\section{System Architecture and C-Band Quantum-Dot Model}\label{sec:model}
\subsection{Architecture and conventions}
Fig.~\ref{fig:arch}(a) shows the geometry of the link and Fig.~\ref{fig:arch}(b) the signal chain. The TX array is prepared in a single-excitation state $\rho_{T}$ and radiates through its own collective coupling; the photon crosses the free-space channel, described by the dyadic Green tensor between TX and RX dots; the RX array absorbs it through its own collective coupling and is left in the spatial state $\rho_{R}$. The RX emission is collected element by element and detected by heterodyne receivers, giving snapshots $\bm{x}[k]$, which feed two separate branches: DOA estimation, returning $(\hat{\theta},\hat{\phi})$, and state reconstruction, returning $\hat{\rho}_{R}$.\\
Quantum operators evolve according to the Schr\"odinger equation, so the physics time convention $e^{-i\omega t}$ is used throughout, and an outgoing spherical wave carries $e^{+ik_{0}R}$. With this convention, the array vector is defined as
\begin{equation}
a_{n}(\shat)=e^{-ik_{0}\shat\cdot\rv_{n}},~
\shat=[\sin\theta\cos\phi,\ \sin\theta\sin\phi,\ \cos\theta]^{T},
\label{eq:a}
\end{equation}
where $\rv_{n}$ is the position of element $n$, $k_{0}=2\pi/\lambda_{0}$ is the free-space wavenumber, $\theta$ and $\phi$ are the elevation (from $+z$) and azimuth angles, and $\shat$ is the corresponding unit vector. The same vector describes two situations: emission toward $\shat$, and reception of a plane wave arriving from a source that lies in the direction $\shat$, whose wave vector is therefore $-k_{0}\shat$. A single vector serving both roles is the array form of reciprocity. A steering state with the opposite phase sign would point toward $-\shat_{0}$, which for a planar array appears as a $180^{\circ}$ azimuth error; all equations and the code use \eqref{eq:a} as written.

\subsection{Effective C-band quantum-dot emitter}
Each dot is treated as a two-level emitter with ground state $|g\rangle$, excited state $|e\rangle$ (the $\hat{\bm{x}}$-polarized exciton $|X\rangle$ in the inset of Fig.~\ref{fig:arch}(a)) and transition angular frequency $\omega_{0}=2\pi c/\lambda_{0}$, where $c$ is the speed of light and $\lambda_{0}=\SI{1550}{\nano\meter}$. This gives $f_{0}=\omega_{0}/2\pi=\SI{193.41}{\tera\hertz}$ and a photon energy $\hbar\omega_{0}=\SI{0.7999}{\electronvolt}$. No transition energy is derived from a confinement model. The material parameters are those measured on droplet-epitaxy InAs/InP dots \cite{anderson2020,wells2023} and are listed with their origin in Table~\ref{tab:param}.\\
All decay rates in this paper are angular-frequency quantities in \si{\per\second}. The radiative decay rate $\Gamma$ is the decay rate of the excited-state population, $\Gamma=1/T_{1}$, where $T_{1}$ is the radiative lifetime. A single dot then emits a Lorentzian spectrum whose full width at half maximum (FWHM) is $\Gamma$ in angular frequency, i.e., $\Gamma/2\pi$ in hertz. The pure-dephasing rate $\gamma_{\phi}$ is defined through the dissipator of Section~\ref{sec:tx}, so that the optical coherence of one dot decays at $T_{2}^{-1}=\Gamma/2+\gamma_{\phi}$ and its emission FWHM becomes $\Gamma+2\gamma_{\phi}=2/T_{2}$. The Fourier limit is $T_{2}=2T_{1}$. The measured Fourier-limited coherence time $2T_{1}=\SI{2.45}{\nano\second}$ \cite{wells2023} gives $T_{1}=\SI{1.225}{\nano\second}$, $\Gamma=\SI{8.163e8}{\per\second}$ ($\Gamma/2\pi=\SI{130}{\mega\hertz}$) and $\hbar\Gamma=\SI{0.537}{\micro\electronvolt}$. The non-resonant coherence time $T_{2}=\SI{447}{\pico\second}$ gives $\gamma_{\phi}=1/T_{2}-\Gamma/2=2.24\Gamma$, and the resonant bound $T_{2}\geq1.5T_{1}$ gives $\gamma_{\phi}\leq(1/1.5-1/2)\Gamma=0.167\Gamma$. Two further linewidths are kept separate from $\Gamma$: $\Gamma_{T}$, the radiative rate of the \emph{TX} dots, varied in Section~\ref{sec:results} to change the photon bandwidth ($\Gamma_{T}=\Gamma$ meaning identical TX and RX dots), and $\gamma$, the FWHM of the transmitted photon spectrum, fixed by the collective TX state and proportional to $\Gamma_{T}$ (further details in Section~\ref{sec:rx}).

\begin{table}[ht!]
\caption{Parameters of the C-Band Quantum-Dot Array Model}
\label{tab:param}
\centering
\scriptsize
\setlength{\tabcolsep}{2.5pt}
\begin{tabular}{@{}p{25mm}lp{24mm}l@{}}
\toprule
Quantity & Symbol & Value & Origin\\
\midrule
Wavelength & $\lambda_{0}$ & \SI{1550}{\nano\meter} & design\\
Frequency & $f_{0}$ & \SI{193.41}{\tera\hertz} & derived\\
Photon energy & $\hbar\omega_{0}$ & \SI{0.7999}{\electronvolt} & derived\\
Wavenumber & $k_{0}$ & \SI{4.054e6}{\per\meter} & derived\\
Radiative lifetime & $T_{1}$ & \SI{1.225}{\nano\second} & from $2T_{1}$ \cite{wells2023}\\
Radiative rate & $\Gamma=1/T_{1}$ & \SI{8.163e8}{\per\second} & derived\\
Fine-struct.\ splitting & FSS & \SI{22.77}{\micro\electronvolt} ($42\hbar\Gamma$) & measured \cite{wells2023}\\
Non-resonant $T_{2}$ & $T_{2}$ & \SI{447}{\pico\second} & measured \cite{wells2023}\\
Dephasing (res.) & $\gamma_{\phi}$ & $\le0.167\,\Gamma$ & $T_{2}\ge1.5T_{1}$ \cite{wells2023}\\
Dephasing (non-res.) & $\gamma_{\phi}$ & $2.24\,\Gamma$ & $1/T_{2}-\Gamma/2$\\
Dipole orientation & $\pv$ & $\hat{\bm{x}}$ (in-plane) & assumed\\
Dipole moment & $\mu_{eg}$ & \SI{3.284e-28}{\coulomb\meter} & from $\Gamma$\\
Elements, spacing & $N$, $d$ & 16 ($4\times4$), $\lambda_{0}/2$ & design\\
Link distance & $R$ & $200\lambda_{0}=\SI{310}{\micro\meter}$ & design\\
Source direction & $(\theta_{s},\phi_{s})$ & $(40^{\circ},30^{\circ})$ & design\\
\bottomrule
\end{tabular}

\vspace{2pt}
\parbox{\columnwidth}{\footnotesize Only $T_{1}$, $T_{2}$ and the FSS are taken from the previously reported measured data; every other emitter quantity follows from them, and the array geometry and link distance are design choices of this study.}
\end{table}

The orientation of the dipoles plays a critical role in quantum dot arrays. The neutral exciton of a self-assembled dot has two orthogonal \emph{in-plane} linear dipoles, split by the fine-structure splitting (FSS), measured as \SI{22.77}{\micro\electronvolt} \cite{wells2023}, which is $42\hbar\Gamma$. A polarization-selective resonant drive therefore addresses one linear dipole, and we take the unit dipole vector $\pv=\hat{\bm{x}}$ for every dot. An out-of-plane dipole, sometimes assumed for convenience, is not the natural orientation of such dots. Another critical factor is \emph{spectral tuning}. As the dots differ in transition energy, so collective effects require tuning to a common frequency, as demonstrated with local strain \cite{grim2019} and electrical control \cite{tiranov2023}. The nominal model assumes tuned, identical dots; residual spectral disorder is a real limitation and is quantified in Section~\ref{sec:robust}.\\
If in a quantum transmit-receive propagation link, the host medium is homogeneous free space, $n_{h}$ becomes $1$. The vacuum-equivalent dipole moment reproducing the measured rate through $\Gamma=\omega_{0}^{3}\mu_{eg}^{2}/(3\pi\varepsilon_{0}\hbar c^{3})$ is $\mu_{eg}=\SI{3.284e-28}{\coulomb\meter}$. Every array result depends on $\Gamma$ and $k_{0}$ only through normalized rates and the products $k_{0}r$. Both arrays are $4\times4$ uniform planar arrays ($N=N_{T}=N_{R}=16$) with element spacing $d=\lambda_{0}/2=\SI{775}{\nano\meter}$ in $x$ and $y$, i.e., a $\SI{2.325}{\micro\meter}\times\SI{2.325}{\micro\meter}$ aperture. The default TX--RX distance is $R=200\lambda_{0}=\SI{310}{\micro\meter}$, and the default source direction seen from the RX is $(\theta_{s},\phi_{s})=(40^{\circ},30^{\circ})$. The distance is a numerical electromagnetic link length chosen to lie well inside the far field of both apertures.

\section{State-Controlled Radiation of the C-Band TX}\label{sec:tx}
\subsection{Collective dynamics from the dyadic Green tensor}
\begin{figure*}[!t]
\centering
\includegraphics[width=0.76\textwidth]{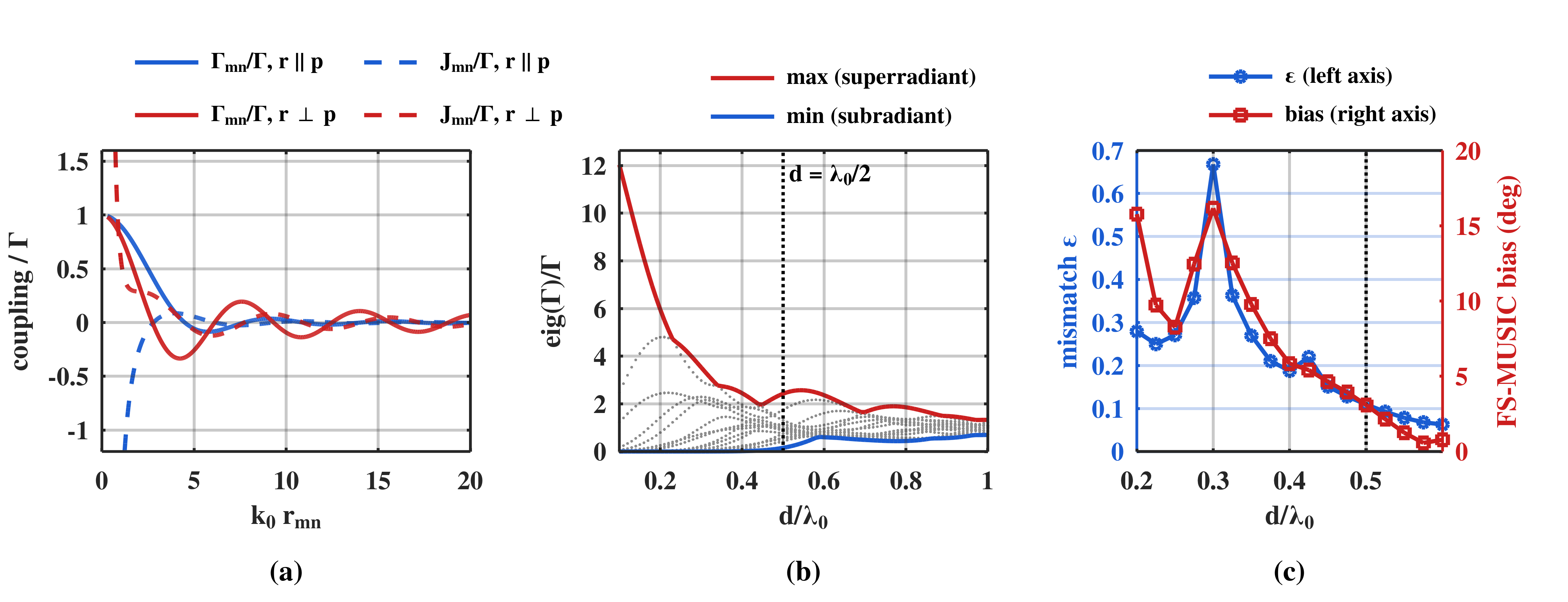}
\caption{Green-tensor coupling of in-plane ($\hat{\bm{x}}$) C-band QD dipoles. (a) $\Gamma_{mn}/\Gamma$ (solid) and $J_{mn}/\Gamma$ (dashed) versus $k_{0}r_{mn}$, parallel and perpendicular to $\pv$. (b) Collective decay rates $\mathrm{eig}(\bm{\Gamma})/\Gamma$ of the $4\times4$ array versus spacing; dotted line: $d=\lambda_{0}/2$. (c) Mismatch $\epsilon_{m}$ between the free-space (FS) and narrowband coupling-aware (CA) manifolds (left axis) and FS-MUSIC bias for a monochromatic source at $(40^{\circ},30^{\circ})$ (right axis).}
\label{fig:coupling}
\end{figure*}

The free-space dyadic Green tensor for the time convention $e^{-i\omega t}$ is
{\small
\begin{align}
\Gt(\rv,\rv')=\frac{e^{ik_{0}R}}{4\pi R}\Big[\Big(1+\frac{i}{k_{0}R}-\frac{1}{k_{0}^{2}R^{2}}\Big)\Id \nonumber\\
+\Big(-1-\frac{3i}{k_{0}R}+\frac{3}{k_{0}^{2}R^{2}}\Big)\hat{\bm{R}}\hat{\bm{R}}\Big],
\label{eq:G}
\end{align}
}
where $\bm{R}=\rv-\rv'$, $R=|\bm{R}|$, $\hat{\bm{R}}=\bm{R}/R$, $\Id$ is the $3\times3$ identity and $\hat{\bm{R}}\hat{\bm{R}}$ the dyadic product, normalized so that $\nabla\times\nabla\times\Gt-k_{0}^{2}\Gt=\Id\,\delta(\bm{R})$. Eliminating the field in the Born--Markov approximation \cite{lehmberg1970,ficek2002} couples two dots $m\neq n$ through \cite{dung2002}
{\small
\begin{align}
J_{mn}-\tfrac{i}{2}\Gamma_{mn}&=-\frac{\mu_{0}\omega_{0}^{2}\mu_{eg}^{2}}{\hbar}\,\pv\cdot\Gt(\rv_{m},\rv_{n})\cdot\pv\nonumber\\
&=-\frac{3\pi\Gamma}{k_{0}}\,\pv\cdot\Gt(\rv_{m},\rv_{n})\cdot\pv ,
\label{eq:JG}
\end{align}
}

where $\mu_{0}$ is the vacuum permeability and $J_{mn}$, $\Gamma_{mn}$ (both in \si{\per\second}) are the coherent (frequency-shift) and dissipative (cross-decay) couplings: $J_{mn}=-(3\pi\Gamma/k_{0})\,\pv\cdot\operatorname{Re}\Gt\cdot\pv$ and $\Gamma_{mn}=(6\pi\Gamma/k_{0})\,\pv\cdot\operatorname{Im}\Gt\cdot\pv$. Since $\operatorname{Im}\Gt(\rv,\rv)=(k_{0}/6\pi)\Id$, the same formula gives $\Gamma_{nn}=\Gamma$, while the divergent $\operatorname{Re}\Gt(\rv,\rv)$ is a Lamb shift already contained in $\omega_{0}$; unequal radiative rates replace $\Gamma$ by $\sqrt{\Gamma_{m}\Gamma_{n}}$. The couplings are collected in the non-Hermitian matrix
{\small
\begin{equation}
\Mm=\bm{\Delta}+\bm{J}-\tfrac{i}{2}\bm{\Gamma},
\label{eq:M}
\end{equation}
}
where $\bm{\Delta}=\operatorname{diag}(\delta_{n})$ contains the detunings $\delta_{n}$ of the dots from $\omega_{0}$, $\bm{J}=[J_{mn}]$ has a zero diagonal, and $\bm{\Gamma}=[\Gamma_{mn}]$ has $\Gamma$ on its diagonal. Throughout the paper the lightface $\Gamma$ is the scalar single-dot decay rate, $\Gamma_{mn}$ its pairwise collective counterpart, and the boldface $\bm{\Gamma}$ the matrix of those entries; the same convention distinguishes $\Gamma_{T}$ (scalar, TX dots) from $\bm{\Gamma}_{TT}$ and $\bm{\Gamma}_{RR}$ (matrix blocks). The eigenvalues $\lambda_{q}$ of $\Mm$ give the collective frequency shifts, $\operatorname{Re}\lambda_{q}$, and half the collective decay rates, $-\operatorname{Im}\lambda_{q}$; Fig.~\ref{fig:coupling} shows the couplings for the arrays used here (further details in Section~\ref{sec:results}-B).

Without a drive the density matrix $\rho$ stays in the span of the collective ground state $|G\rangle$ and the single-excitation states $|e_{n}\rangle$ (dot $n$ excited, all others in $|g\rangle$). Its excited-state block $\Cm(t)$, with elements $C_{mn}=\langle e_{m}|\rho|e_{n}\rangle$, obeys
\begin{equation}
\dot{\Cm}=-i\big(\Mm\Cm-\Cm\Mm^{H}\big)-2\gamma_{\phi}\big(\Cm-\operatorname{diag}\Cm\big),
\label{eq:Cdot}
\end{equation}
where $(\cdot)^{H}$ is the conjugate transpose and $\operatorname{diag}\Cm$ keeps only the diagonal of $\Cm$. The last term is the pure-dephasing dissipator $\mathcal{D}[\rho]=(\gamma_{\phi}/2)\sum_{n}(\hat{\sigma}^{z}_{n}\rho\hat{\sigma}^{z}_{n}-\rho)$ restricted to one excitation. For one dot it damps the optical coherence $\langle e|\rho|g\rangle$ at $\gamma_{\phi}$, giving $T_{2}^{-1}=\Gamma/2+\gamma_{\phi}$ with radiative decay, the relation used in Table~\ref{tab:param}; between two dots each $\hat{\sigma}^{z}$ factor contributes $\gamma_{\phi}$, hence $2\gamma_{\phi}$ in \eqref{eq:Cdot}. For $\gamma_{\phi}=0$ the solution is $\Cm(t)=e^{-i\Mm t}\Cm(0)e^{i\Mm^{H}t}$. With the eigen-decomposition $\Mm=\bm{V}\bm{\Lambda}\bm{V}^{-1}$, $\bm{\Lambda}=\operatorname{diag}(\lambda_{k})$, the time integral that governs every energy-type quantity below has the closed form
{\small
\begin{equation}
\bm{X}=\!\int_{0}^{\infty}\!\Cm(t)\,dt=\bm{V}\Big[\frac{\big(\bm{V}^{-1}\Cm(0)\bm{V}^{-H}\big)_{kl}}{i(\lambda_{k}-\lambda_{l}^{*})}\Big]\bm{V}^{H},
\label{eq:X}
\end{equation}
}

where the bracket denotes the matrix of the $(k,l)$-th element. The denominator never vanishes, because every collective mode decays ($\operatorname{Im}\lambda_{k}<0$). For $\gamma_{\phi}>0$, \eqref{eq:Cdot} is a linear system $\tfrac{d}{dt}\operatorname{vec}\Cm=\mathcal{L}\operatorname{vec}\Cm$, and $\operatorname{vec}\bm{X}=-\mathcal{L}^{-1}\operatorname{vec}\Cm(0)$, where $\operatorname{vec}$ stacks columns and $\mathcal{L}$ is the Liouvillian matrix of \eqref{eq:Cdot}.

\subsection{Far-field pattern and the element factor}
Far from the array, the positive-frequency field radiated toward $\shat$ is proportional to $(\Id-\shat\shat)\cdot\pv\sum_{n}a_{n}(\shat)\hat{\sigma}_{n}^{-}$, where $(\Id-\shat\shat)$ projects onto the plane transverse to $\shat$. The first-order (intensity) pattern is therefore
\begin{equation}
G^{(1)}(\shat,t)=\mathrm{EF}(\shat)\,\av^{H}(\shat)\RQ(t)\av(\shat),
\label{eq:G1}
\end{equation}
where $\mathrm{EF}(\shat)=1-(\pv\cdot\shat)^{2}$ is the element factor of a $\pv$-directed dipole and $[\RQ]_{mn}=\langle\hat{\sigma}^{+}_{m}\hat{\sigma}^{-}_{n}\rangle$. In the single-excitation manifold $\hat{\sigma}^{+}_{m}\hat{\sigma}^{-}_{n}=|e_{m}\rangle\langle e_{n}|$, so $\langle\hat{\sigma}^{+}_{m}\hat{\sigma}^{-}_{n}\rangle=\langle e_{n}|\rho|e_{m}\rangle=C_{nm}$; that is, $\RQ=\Cm^{T}$. The transpose is not an error: it reflects the two index orders used for the same matrix, operator order for $\RQ$ and ket--bra order for $\Cm$. The element factor is not a free modeling choice. Integrating \eqref{eq:G1} over the sphere and using the plane-wave representation $\operatorname{Im}\Gt(\bm{R})=(k_{0}/16\pi^{2})\oint(\Id-\shat\shat)e^{ik_{0}\shat\cdot\bm{R}}d\Omega$ gives the sum rule
\begin{equation}
\frac{3}{8\pi}\oint\mathrm{EF}(\shat)\,\av(\shat)\av^{H}(\shat)\,d\Omega=\frac{\bm{\Gamma}}{\Gamma} ,
\label{eq:sumrule}
\end{equation}
where $d\Omega$ is the element of solid angle and the right-hand side is the collective decay \emph{matrix} $\bm{\Gamma}=[\Gamma_{mn}]$ divided by the scalar single-dot rate $\Gamma$. Hence $(3\Gamma/8\pi)\oint G^{(1)}d\Omega=\Tr(\bm{\Gamma}\RQ)=-\tfrac{d}{dt}\Tr\Cm$ for $\gamma_{\phi}=0$: the photon flux through a large sphere equals the loss of excited population only if the element factor belongs to the dipole that enters $\bm{\Gamma}$, which isotropic elements with dipole coupling do not.\\
Two patterns are reported in the next. The \emph{prepared-state} pattern from $\RQ(0)$, and the \emph{emitted-photon} pattern from $\bm{X}^{T}$ of \eqref{eq:X}, the angular distribution of the complete photon. They differ only because the collective modes in the prepared state decay at different rates. The directivity is $D_{0}=4\pi\max_{\shat}G^{(1)}/\oint G^{(1)}d\Omega$; we write $D_{0}$ for the prepared state and $D_{0}^{\mathrm{int}}$ for the emitted photon.

\subsection{State--geometry factorization and planar mirror symmetry}
Equation \eqref{eq:G1} factorizes the pattern: $\av(\shat)$ and the element factor are fixed by the positions and the dipole orientation, $\RQ$ by the state, whose diagonal $[\RQ]_{nn}$ is the excitation probability of dot $n$ and whose off-diagonal entries carry the mutual coherence of two dots and their relative optical phase. Reconfiguring the array means changing $\RQ$ while aperture, spacing, and elements stay untouched: such an array is state-programmable rather than feed-programmable.

The factorization also fixes what state control cannot do. For a planar array in $z=0$, $\shat\cdot\rv_{n}$ depends on $\theta$ only through $\sin\theta$, and for an in-plane dipole $\mathrm{EF}$ depends on $\theta$ the same way, so both factors of \eqref{eq:G1} are invariant under $\theta\rightarrow180^{\circ}-\theta$:
\begin{equation}
G^{(1)}(\theta,\phi)=G^{(1)}(180^{\circ}-\theta,\phi)\quad\text{for every state }\rho .
\label{eq:mirror}
\end{equation}
numerically, the largest relative asymmetry over the steering states of Section~\ref{sec:results}-C is $1.3\times10^{-15}$. Every steered beam therefore carries a mirror lobe at the same azimuth in the lower half space that no choice of $\RQ$ removes, so single-hemisphere operation needs a ground plane or a directive element, as for a classical planar array, and this symmetry -- not an estimator heuristic -- is why the DOA search of Section~\ref{sec:doa} is restricted to the upper half space. Asymmetry in azimuth, where the array has extent, remains state-controlled and is measured by $\eta_{A}$ (Section~\ref{sec:results}-C); it is a pattern property only, since reciprocity constrains the transmit--receive relation and not the spatial symmetry of a pattern.

\subsection{Steering by state preparation and classical equivalence}
\begin{figure}[!t]
\centering
\includegraphics[width=\columnwidth]{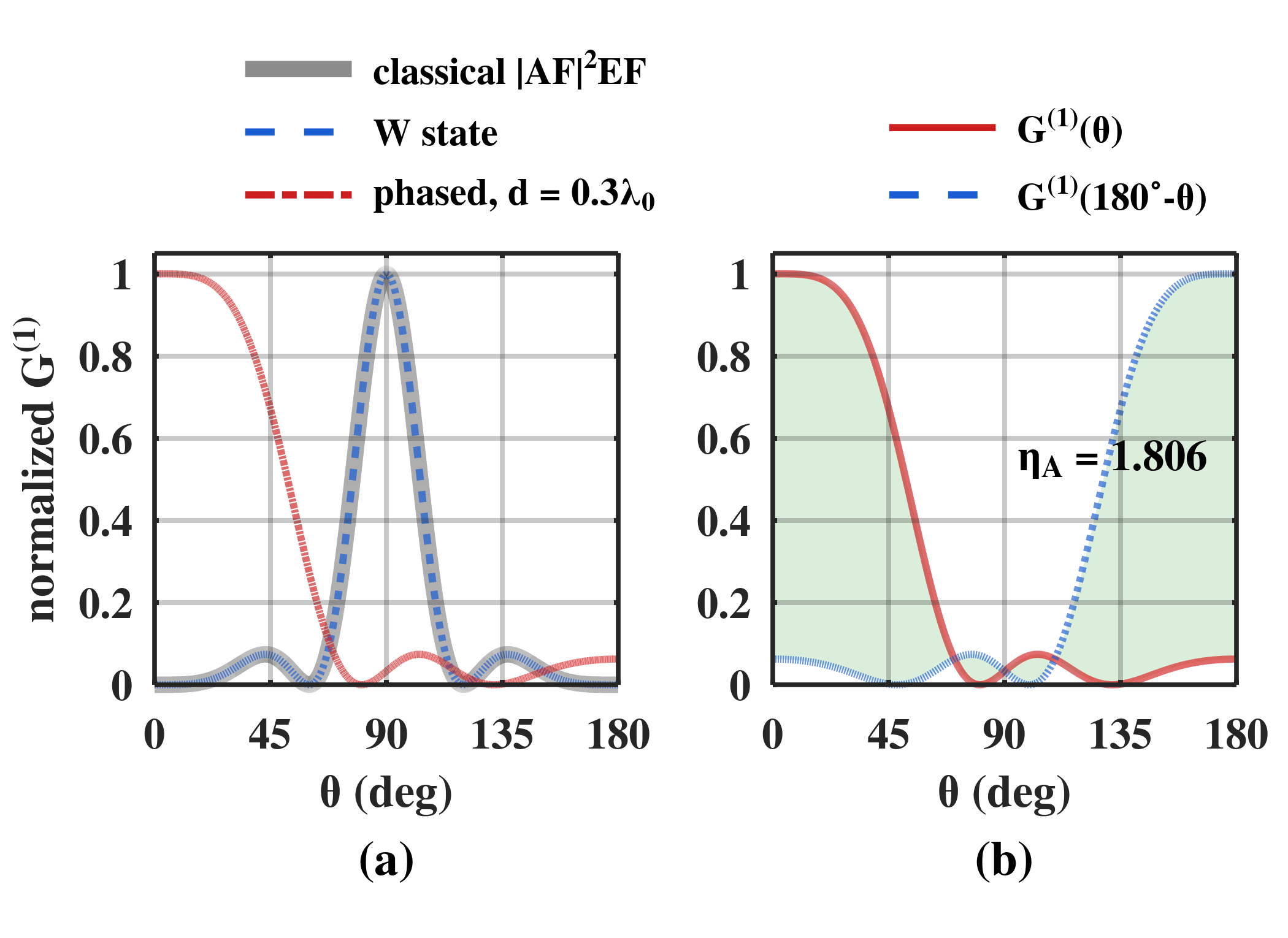}
\caption{Four-element linear array (along $z$, $\hat{\bm{x}}$ dipoles), $yz$-plane cut. (a) The classical $|\mathrm{AF}|^{2}\mathrm{EF}$ pattern (thick grey) and the $W$ state (dashed) coincide; the phased state at $d=0.3\lambda_{0}$ (dash-dot) radiates toward endfire. (b) Forward and mirrored patterns of the phased state; the shaded area is the asymmetry that defines $\eta_{A}$.}
\label{fig:linear}
\end{figure}

\begin{figure*}[!t]
\centering
\includegraphics[width=0.76\textwidth]{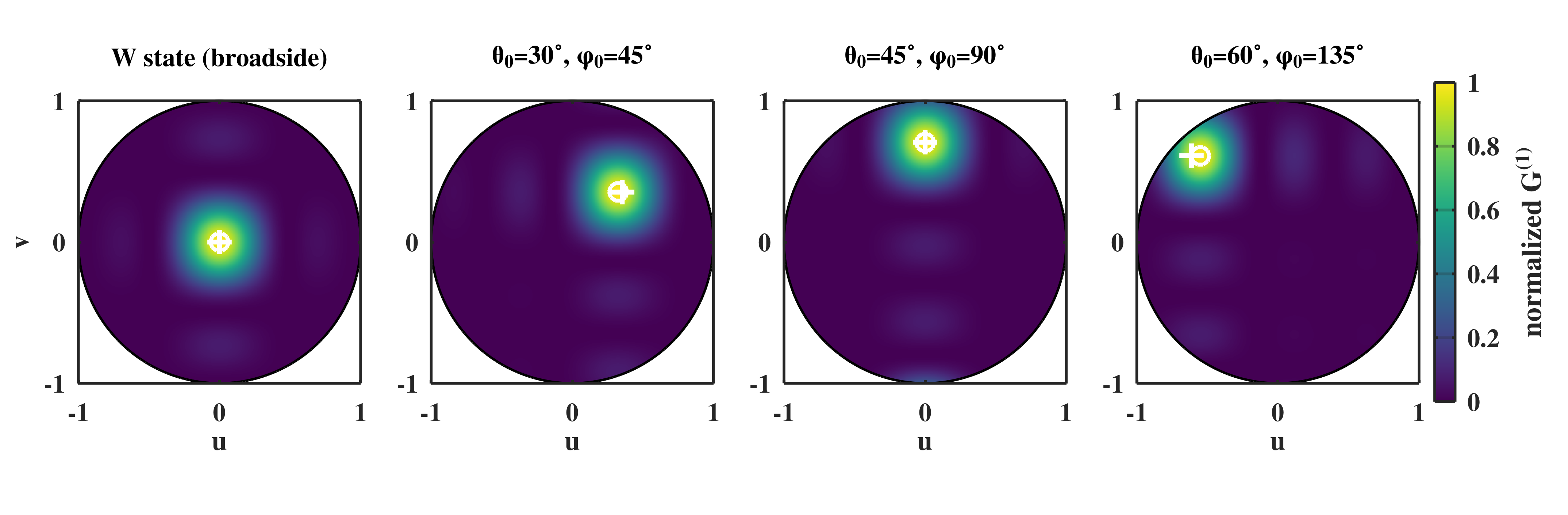}
\caption{Normalized patterns $G^{(1)}$ of the prepared states of the $4\times4$ array on the disc $(u,v)=(\sin\theta\cos\phi,\sin\theta\sin\phi)$, the projection of $\shat$ onto the array plane. From left: $W$ state and timed Dicke states with commanded $(\theta_{0},\phi_{0})$. White cross: commanded $\shat_{0}$; white circle: actual peak.}
\label{fig:patterns}
\end{figure*}

\begin{figure*}[!t]
\centering
\includegraphics[width=0.76\textwidth]{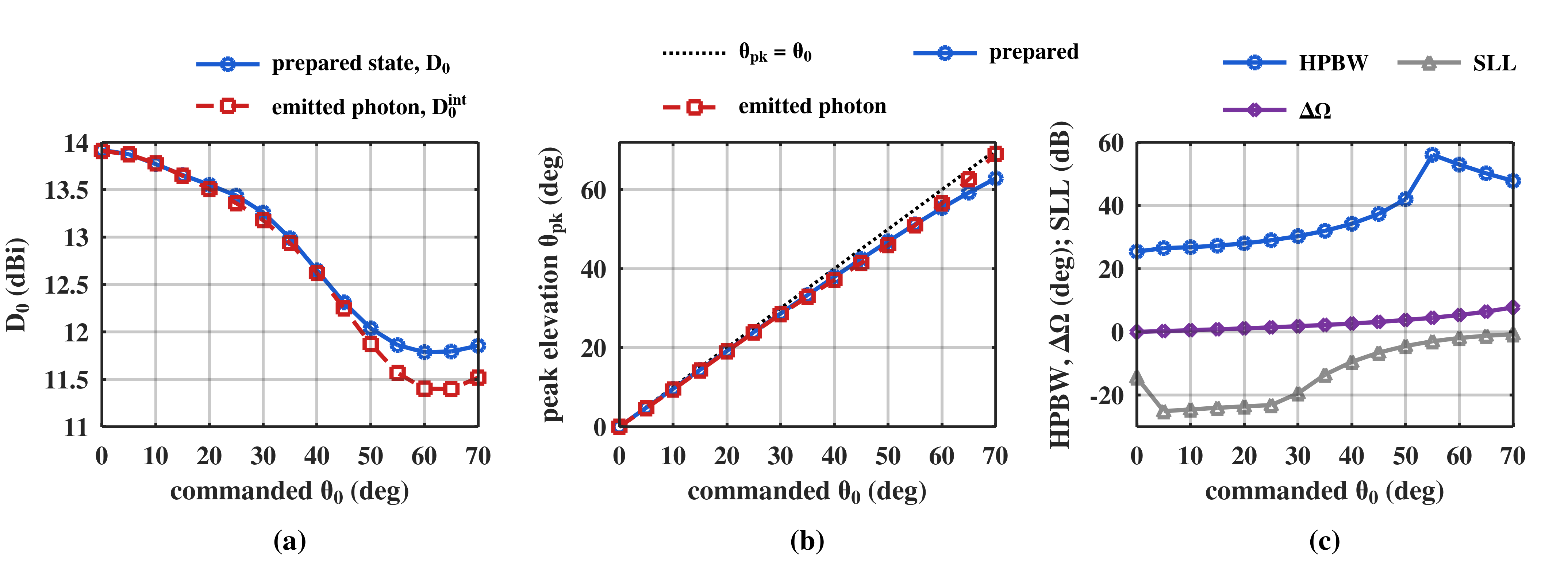}
\caption{Scan performance at $\phi_{0}=45^{\circ}$. (a) Directivity of the prepared state, $D_{0}$, and of the emitted photon, $D_{0}^{\mathrm{int}}$. (b) Peak versus commanded elevation; dotted line: $\theta_{\mathrm{pk}}=\theta_{0}$. (c) HPBW, SLL and pointing displacement $\Delta\Omega$.}
\label{fig:scan}
\end{figure*}

Following \cite{scully2006} and \cite{svidzinsky2010}, the timed Dicke state  
\begin{equation}
|\psi_{1}(\shat_{0})\rangle=\sum_{n}c_{n}|e_{n}\rangle,\quad c_{n}=\frac{a_{n}^{*}(\shat_{0})}{\sqrt{N}}=\frac{e^{ik_{0}\shat_{0}\cdot\rv_{n}}}{\sqrt{N}},
\label{eq:TD}
\end{equation}
with commanded direction $\shat_{0}$ and $\Cm(0)=\cv\cv^{H}$, gives $G^{(1)}(\shat,0)=\mathrm{EF}(\shat)|\sum_{n}c_{n}a_{n}(\shat)|^{2}$, exactly the pattern of a classical phased array of $\pv$-directed dipoles with weights $w_{n}=c_{n}$. The equivalence holds for every pure single-excitation state and settles what can be claimed: first-order radiation of a state-controlled array is steered, shaped and limited as a classical array is. Two consequences follow. The array factor peaks at $\shat_{0}$ but its product with $\mathrm{EF}$ does not, so for $\pv=\hat{\bm{x}}$ the peak is displaced toward broadside and the $yz$ plane; we call this great-circle offset the element-factor pointing displacement $\Delta\Omega$, a property of the element rather than an error of the state. And \eqref{eq:TD} is not an eigenmode of $\Mm$: it decays at $\Gamma_{\mathrm{st}}=\cv^{H}\bm{\Gamma}\cv$ and is shifted by $\cv^{H}\bm{J}\cv$ only at $t=0$, so the emitted-photon pattern departs from the prepared one. Figs.~\ref{fig:linear}--\ref{fig:scan} illustrate both points numerically (Section~\ref{sec:results}-C).

\subsection{Second-order coherence as the quantum discriminator}
\begin{figure*}[!t]
\centering
\includegraphics[width=0.76\textwidth]{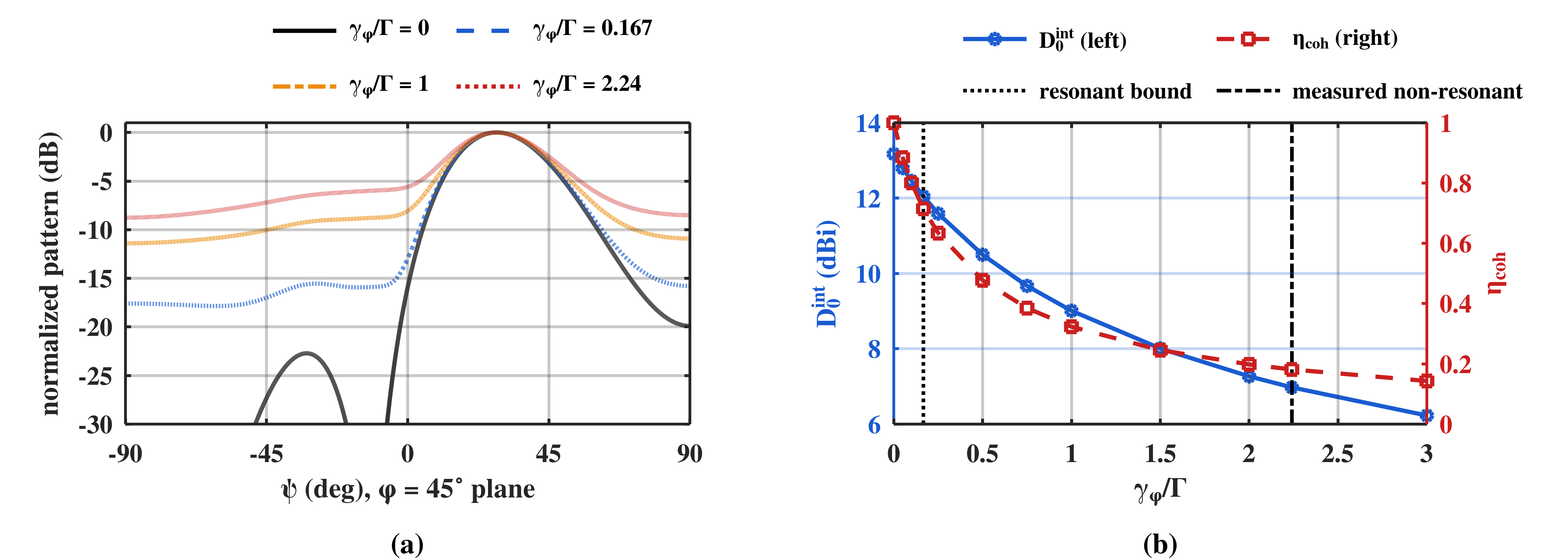}
\caption{Pure dephasing of the steered transmitter; all curves are emitted-photon (time-integrated) patterns of the $(30^{\circ},45^{\circ})$ state. (a) Cuts in the plane containing the azimuths $\phi=45^{\circ}$ and $\phi=225^{\circ}$, plotted against the signed elevation $\psi$, which equals $+\theta$ in the $\phi=45^{\circ}$ half-plane and $-\theta$ in the $\phi=225^{\circ}$ half-plane. (b) $D_{0}^{\mathrm{int}}$ (left) and coherence ratio $\eta_{\mathrm{coh}}$ (right); the measured resonant bound and non-resonant value are marked.}
\label{fig:txdeph}
\end{figure*}

A classical phased array cannot reproduce the photon statistics. With $\hat{A}(\shat)=\sum_{n}a_{n}(\shat)\hat{\sigma}^{-}_{n}$ and the shorthand $\hat{A}_{i}=\hat{A}(\shat_{i})$, $\mathrm{EF}_{i}=\mathrm{EF}(\shat_{i})$,
\begin{equation}
G^{(2)}(\shat_{1},\shat_{2})=\mathrm{EF}_{1}\mathrm{EF}_{2}\,\langle\hat{A}^{\dagger}_{1}\hat{A}^{\dagger}_{2}\hat{A}_{2}\hat{A}_{1}\rangle ,
\label{eq:G2}
\end{equation}
which vanishes identically on the single-excitation manifold, two lowering operators annihilating any state with one excitation. Real sources contain a small two-excitation admixture, modelled as $\rho=(1-p_{2})|\psi_{1}\rangle\langle\psi_{1}|+p_{2}|\psi_{2}\rangle\langle\psi_{2}|$ with two-excitation probability $p_{2}$ and $|\psi_{2}\rangle\propto\sum_{m<n}c_{m}c_{n}|e_{m}e_{n}\rangle$; in the commanded direction this gives $G^{(1)}\propto(1-p_{2})N+2p_{2}(N-1)$ and $G^{(2)}\propto2p_{2}N(N-1)$, so $g^{(2)}=G^{(2)}/(G^{(1)})^{2}$ is
\begin{equation}
g^{(2)}(\shat_{0})=\frac{2p_{2}N(N-1)}{\big[(1-p_{2})N+2p_{2}(N-1)\big]^{2}}\approx\frac{2p_{2}(N-1)}{N}.
\label{eq:g2}
\end{equation}
The coherent-state reference has $g^{(2)}=1$ and thermal light $g^{(2)}=2$; Table~\ref{tab:g2} lists \eqref{eq:g2} computed from the full operator algebra, which agrees with the closed form to all printed digits. These statistics identify the source as a single-photon emitter and justify the one-photon measurement model of Section~\ref{sec:meas}, but, as Section~\ref{sec:rx} shows, they are not the limiting factor for the receiver. Dephasing, by contrast, converts part of the coherent steered emission into incoherent element-like emission (Fig.~\ref{fig:txdeph}).

\begin{table}[!t]
\caption{$g^{(2)}$ in the Commanded Direction With a Two-Excitation Admixture $p_{2}$ (Steering State $(30^{\circ},45^{\circ})$)}
\label{tab:g2}
\centering
\footnotesize
\begin{tabular}{@{}lccccc@{}}
\toprule
$p_{2}$ & 0 & 0.001 & 0.01 & 0.05 & 0.1\\
\midrule
$g^{(2)}$, $N=4$ & 0 & 0.0015 & 0.0149 & 0.0714 & 0.1361\\
$g^{(2)}$, $N=16$ & 0 & 0.0019 & 0.0184 & 0.0861 & 0.1585\\
\midrule
Coherent-state reference & \multicolumn{5}{c}{$g^{(2)}=1$ for all $p_{2}$}\\
\bottomrule
\end{tabular}
\end{table}

\section{TX-RX Channel Modeling Using Dyadic Green-Function Formalism}\label{sec:channel}
\subsection{Block formulation of transmitter, channel and receiver}
The RX array lies in the plane $z=0$, centred at the origin, and the TX array is a parallel copy centred at $R\shat_{s}$, where $\shat_{s}$ is the source direction seen from the RX. The positions are $\rv^{T}_{n}$ ($n=1,\dots,N_{T}$) and $\rv^{R}_{m}$ ($m=1,\dots,N_{R}$). Stacking the TX amplitudes above the RX amplitudes and applying \eqref{eq:JG} to all $N_{T}+N_{R}=32$ dots gives
\begin{gather}
\Mm=\begin{bmatrix}\Mm_{TT}&\Mm_{TR}\\ \Mm_{RT}&\Mm_{RR}\end{bmatrix},\nonumber\\
[\Mm_{XY}]_{mn}=-\frac{3\pi\Gamma}{k_{0}}\,\pv\cdot\Gt(\rv^{X}_{m},\rv^{Y}_{n})\cdot\pv ,\quad X,Y\in\{T,R\},
\label{eq:Mblock}
\end{gather}
for every pair of distinct dots, with the diagonal entries $\delta_{n}-i\Gamma/2$ of \eqref{eq:M} on the diagonals of $\Mm_{TT}$ and $\Mm_{RR}$. The first index of a block refers to the dot that is acted on, the second to the dot that acts; hence $\Mm_{RT}$ carries excitation from TX to RX and $\Mm_{TR}=\Mm_{RT}^{T}$ by reciprocity. We write $\bm{G}_{XY}=[\pv\cdot\Gt(\rv^{X}_{m},\rv^{Y}_{n})\cdot\pv]$ for the corresponding Green blocks. $\bm{G}_{TT}$ shapes the emitted photon (Section~\ref{sec:tx}), $\bm{G}_{RT}$ is the propagation channel, and $\bm{G}_{RR}$ distorts the receive manifold (Section~\ref{sec:rx}). The TX is prepared in the steering state that points back at the receiver. Since $\shat_{s}$ is defined from the RX toward the TX, the photon must propagate toward $-\shat_{s}$, so the state phase of \eqref{eq:TD} is taken with $\shat_{0}=-\shat_{s}$ and the sign of the exponent is opposite to that in \eqref{eq:TD}: $c_{n}=e^{-ik_{0}\shat_{s}\cdot\rv^{T}_{n}}/\sqrt{N_{T}}$. The RX starts in its ground state, and
\begin{gather}
\Cm(0)=\begin{bmatrix}\cv\cv^{H}&\bm{0}\\ \bm{0}&\bm{0}\end{bmatrix},~
\Cm_{R}=\bm{X}_{RR},~ P_{\mathrm{abs}}=\Tr\big(\bm{\Gamma}_{RR}\Cm_{R}\big),
\label{eq:link}
\end{gather}
where $\bm{X}_{RR}$ is the lower-right $N_{R}\times N_{R}$ block of \eqref{eq:X} evaluated for the full $32\times32$ matrix $\Mm$, and $\bm{\Gamma}_{RR}$ is the RX block of $\bm{\Gamma}$. Because all excitation that enters the RX is eventually radiated by it, $P_{\mathrm{abs}}$ is the probability that the photon is absorbed by the RX array. Unlike a semiclassical receiver model, this formulation contains no expectation value of the field operator, so it applies to a Fock state \cite{glauber1963}; it conserves probability photon by photon, including RX-to-TX back-action through $\Mm_{TR}$; and the full spatial near-field terms of the dyadic Green tensor are retained in every TX--RX pair.\\
To understand what the above Markov elimination retains and neglects,
it keeps the full spatial dependence of $\Gt$ between every pair of dots, including the retardation phase $e^{ik_{0}R}$ and all $1/R$, $1/R^{2}$ and $1/R^{3}$ terms, but neglects the propagation delay $\tau=R/c$ in the \emph{envelopes}, replacing $c_{n}(t-\tau)$ by $c_{n}(t)$, and evaluates $\Gt$ at $\omega_{0}$ rather than across the photon band. The neglected delay and the single-frequency evaluation both give errors of relative order $\Gamma\tau=\Gamma R/c$, so the model requires $\Gamma\tau=\Gamma R/c\ll1$; this quantity is $8.4\times10^{-4}$ at the default $R=\SI{310}{\micro\meter}$ and $4.2\times10^{-3}$ at the largest distance studied, $1000\lambda_{0}$. Once $\Gamma\tau$ approaches $0.1$ ($R$ of the order of centimetres) the delayed amplitudes must be kept and the link treated as a cascaded open system \cite{gardiner1985,carmichael1993}, the RX--TX back-action then arriving after the round-trip delay $2\tau$.

\begin{figure*}[!t]
\centering
\includegraphics[width=0.76\textwidth]{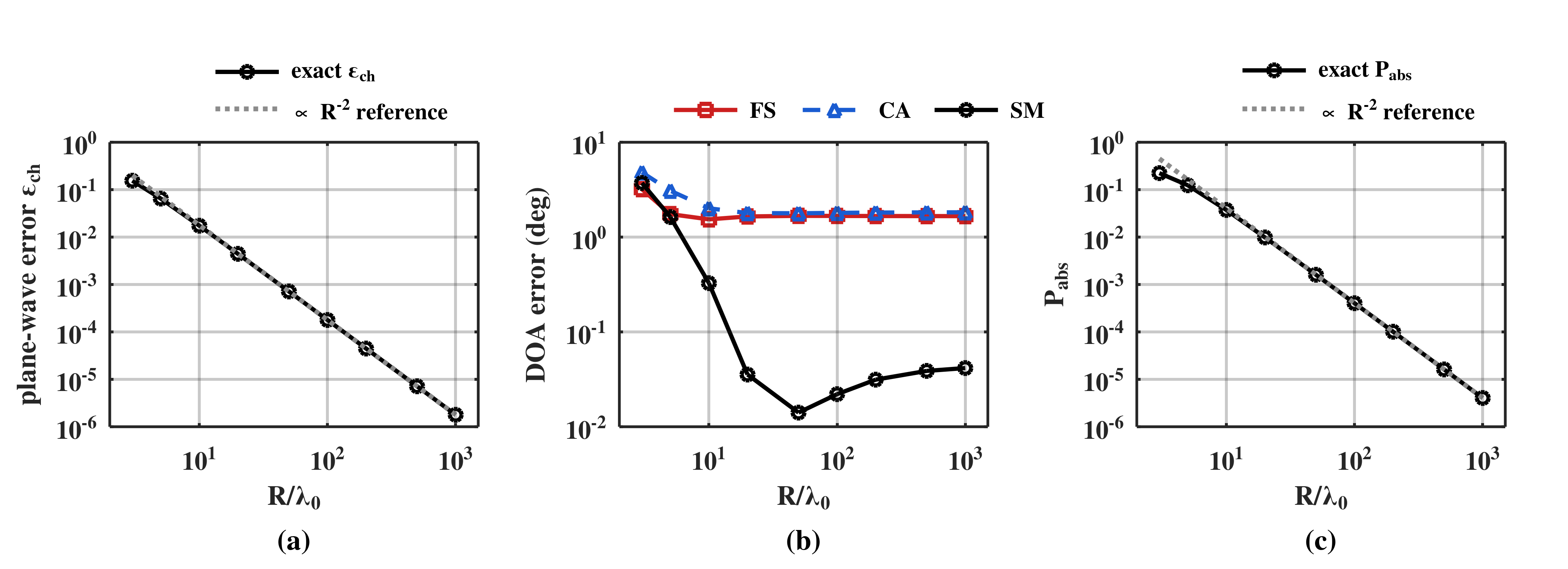}
\caption{Single-photon link versus TX--RX distance for a source at $(40^{\circ},30^{\circ})$; the distances are numerical electromagnetic link lengths, not demonstrated communication ranges. (a) Plane-wave channel error $\epsilon_{\mathrm{ch}}$ with an $R^{-2}$ reference line. (b) MUSIC error with the free-space (FS), narrowband coupling-aware (CA) and spectrally matched (SM) manifolds, computed from the exact covariance at $P_{d}=0.1$. (c) Absorption probability $P_{\mathrm{abs}}$ of the RX array with an $R^{-2}$ reference line.}
\label{fig:link}
\end{figure*}

\subsection{Near- and far-field forms of the channel}
The field that reaches the RX dots at frequency $\omega$ is represented by the vector $\bm{h}(\omega)=\Mm_{RT}\,\cv_{T}(\omega)$, in \si{\per\second}, where $\cv_{T}(\omega)$ is the Fourier transform of the TX amplitudes defined in Section~\ref{sec:rx}. For $R$ well beyond the Fraunhofer distance of the apertures, $|\rv^{R}_{m}-\rv^{T}_{n}|\approx R-\shat_{s}\cdot\rv^{R}_{m}+\shat_{s}\cdot(\rv^{T}_{n}-R\shat_{s})$, and \eqref{eq:G} and \eqref{eq:Mblock} provide
\begin{subequations}   
\begin{align}
\bm{h}&\approx\kappa\,\av(\shat_{s}),\\
\kappa&=-\frac{3\Gamma e^{ik_{0}R}}{4k_{0}R}\,\mathrm{EF}(\shat_{s})\sum_{n}e^{ik_{0}\shat_{s}\cdot(\rv^{T}_{n}-R\shat_{s})}c_{n},
\label{eq:hfar}
\end{align}
\end{subequations}
where the scalar $\kappa$ contains the path loss, the element factors and the TX array factor. For the steering state the sum equals $\sqrt{N_{T}}$ times a unit-modulus phase. The far-field channel is therefore rank one, its only angular content being \eqref{eq:a}, and the departure from \eqref{eq:hfar} is measured by the plane-wave error
\begin{equation}
\epsilon_{\mathrm{ch}}=1-\frac{|\av^{H}\bm{h}|^{2}}{\|\av\|^{2}\|\bm{h}\|^{2}},
\label{eq:epsch}
\end{equation}
which scales as $(L_{\mathrm{ap}}^{2}/\lambda_{0}R)^{2}\propto R^{-2}$ for an aperture of largest dimension $L_{\mathrm{ap}}$; here $L_{\mathrm{ap}}=\sqrt{2}\times\SI{2.325}{\micro\meter}$ and the Fraunhofer distance is \SI{14}{\micro\meter}, or $9\lambda_{0}$. Since $|\kappa|^{2}\propto R^{-2}$, $P_{\mathrm{abs}}$ also decays as $R^{-2}$; this is the free-space Friis law applied to one photon. Fig.~\ref{fig:link} confirms both of these $R^{-2}$ scalings numerically.

\section{Collectively Coupled RX}\label{sec:rx}
\subsection{Single-photon response of the receiver}
For a pure state with $\gamma_{\phi}=0$, \eqref{eq:Cdot} is equivalent to the amplitude equation $i\dot{\cv}=\Mm\cv$ with $\cv=[\cv_{T};\cv_{R}]$ and $\Cm=\cv\cv^{H}$. We define the transform $\cv(\omega)=\int_{0}^{\infty}\cv(t)e^{i\omega t}dt$, with $\omega$ measured from $\omega_{0}$. With $\cv_{R}(0)=\bm{0}$ the RX block of the amplitude equation gives
\begin{align}
\cv_{R}(\omega)&=(\omega\Id-\Mm_{RR})^{-1}\Mm_{RT}\,\cv_{T}(\omega)\nonumber\\
&=(\omega\Id-\Mm_{RR})^{-1}\bm{h}(\omega),
\label{eq:cR}
\end{align}
where $\Id$ is now the $N_{R}\times N_{R}$ identity. Equation \eqref{eq:cR}, written with the full excitation vector $\bm{h}(\omega)=\Mm_{RT}\cv_{T}(\omega)$, is exact within the present model; it is not restricted to a plane-wave excitation. In the far-field regime, where the TX-to-RX field factorizes as $\bm{h}(\omega)\approx\kappa(\omega)\av(\shat)$ by \eqref{eq:hfar}, the response reduces to the coupling-aware form $\cv_{R}(\omega)\propto\kappa(\omega)(\omega\Id-\Mm_{RR})^{-1}\av(\shat)$, which inherits the far-field approximation and should not be read as exact in the general near-field case. The TX amplitudes follow from $\cv_{T}(\omega)=i(\omega\Id-\Mm_{TT})^{-1}\cv_{T}(0)$ up to the back-action term $\Mm_{TR}\cv_{R}$, of relative order $(k_{0}R)^{-1}$, which the numerics retain in any case. The usual semiclassical receiver is recovered as follows. A weak coherent field of frequency $\omega$, described by the Rabi-frequency vector $\bm{\Omega}$ at the RX dots, produces in linear response the steady amplitudes $\bm{\beta}(\omega)=\tfrac{1}{2}(\omega\Id-\Mm_{RR})^{-1}\bm{\Omega}$, i.e., $\bm{\beta}=-\tfrac{1}{2}\Mm_{R}^{-1}\bm{\Omega}$ with $\Mm_{R}=\Mm_{RR}-\omega\Id$.

\begin{proposition}\label{prop:1}
Let the field incident on the RX be radiated by the same TX dipoles, either as a single photon or as a weak coherent field. Then, at every frequency $\omega$, the RX single-excitation amplitude vector $\cv_{R}(\omega)$ of \eqref{eq:cR} is parallel to the linear-response vector $\bm{\beta}(\omega)$. The equivalence concerns only the spatial shape of the amplitude response at each frequency, so, in the far-field regime where $\bm{h}(\omega)\approx\kappa(\omega)\av(\shat)$, the coupling-aware form $(\omega\Id-\Mm_{RR})^{-1}\av$ gives the exact frequency-by-frequency spatial response of the reduced plane-wave model, whereas a single frequency-independent CA manifold evaluated at one $\omega_{c}$ is exact only in the narrowband limit. It does not state that the two excitations are the same physical state: they differ in their spectral weight $|\bm{h}(\omega)|^{2}$, in their photon-number distribution and in all higher-order correlations.
\end{proposition}
\begin{IEEEproof}
Both $\bm{h}(\omega)$ and $\bm{\Omega}$ are the field of the TX polarization carried to the RX dipoles by the same Green block: $\bm{h}=\Mm_{RT}\cv_{T}(\omega)$ and $\bm{\Omega}\propto\Mm_{RT}\bm{w}$, where $\bm{w}\propto\cv_{T}(\omega)$ are the classical dipole currents that radiate the same field. The two vectors are then mapped to the RX amplitudes by the same resolvent $(\omega\Id-\Mm_{RR})^{-1}$. The single-excitation equations are linear, so no approximation on the field state enters \eqref{eq:cR}.
\end{IEEEproof}

Proposition~\ref{prop:1} answers the question raised in Section~\ref{sec:intro}. In the far-field regime the coupling-aware manifold $\Mm_{R}^{-1}\av$ remains valid under single-photon excitation, but only frequency by frequency. The receiver therefore responds to the \emph{spectrum} of the photon, and that spectrum is set by the transmitter.

\subsection{Receiver state and its mixedness}
\begin{figure*}[!t]
\centering
\includegraphics[width=0.76\textwidth]{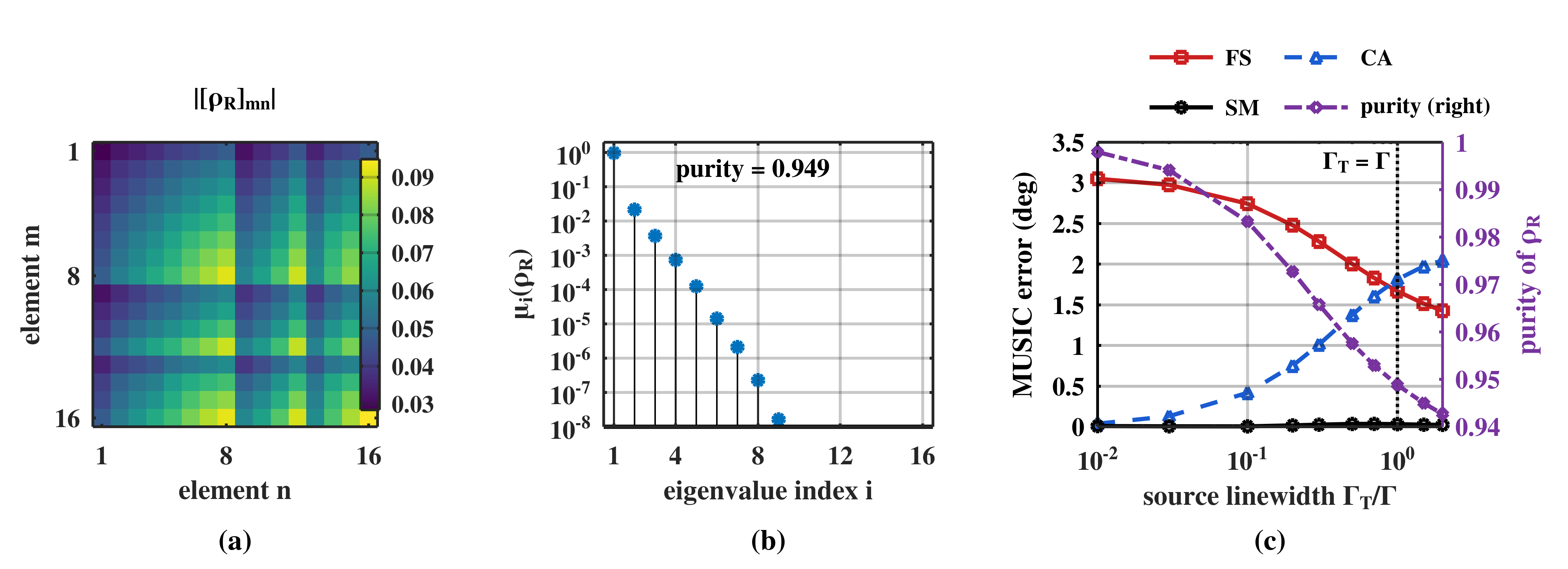}
\caption{Receiver state $\rho_{R}$ (spatial state of the absorbed photon, time traced out) for the default link. (a) $|[\rho_{R}]_{mn}|$. (b) Eigenvalues $\mu_{i}$. (c) MUSIC error with the FS, CA and SM manifolds (left axis) and purity $\Tr\rho_{R}^{2}$ (right axis) versus the TX radiative rate $\Gamma_{T}$, which sets the photon linewidth; $\Gamma_{T}=\Gamma$ (dotted) is the lifetime-limited photon, $\Gamma_{T}\rightarrow0$ the narrowband limit of Proposition~\ref{prop:1}.}
\label{fig:rxstate}
\end{figure*}

By Parseval's theorem the time-integrated RX excitation is
{\small
\begin{align}
\Cm_{R}&=\int_{0}^{\infty}\cv_{R}(t)\cv_{R}^{H}(t)\,dt\nonumber\\
&=\frac{1}{2\pi}\int_{-\infty}^{\infty}(\omega\Id-\Mm_{RR})^{-1}\bm{h}(\omega)\nonumber\\
&\qquad\times\bm{h}^{H}(\omega)(\omega\Id-\Mm_{RR})^{-H}d\omega .
\label{eq:CRparseval}
\end{align}
}
We define the \emph{receiver state} as the unit-trace matrix
\begin{equation}
\rho_{R}=\Cm_{R}/\Tr\Cm_{R}.
\label{eq:rhoR}
\end{equation}
If each RX dot radiates at the same rate into its own collection channel (Section~\ref{sec:meas}), the photon leaving the RX is in the state $|\psi_{\mathrm{out}}\rangle\propto\int dt\sum_{n}c_{R,n}(t)|1_{n}(t)\rangle$, with $|1_{n}(t)\rangle$ one photon in channel $n$ at time $t$; tracing out the time (equivalently frequency) degree of freedom leaves $\int\cv_{R}\cv_{R}^{H}dt/\Tr(\cdot)=\rho_{R}$. Thus $\rho_{R}$ is the spatial state of the absorbed and re-emitted photon, conditioned on absorption, and the symbol is used only in this sense.\\
Although the joint TX--field--RX state is pure, $\rho_{R}$ is mixed whenever $(\omega\Id-\Mm_{RR})^{-1}\bm{h}(\omega)$ turns in direction across the band occupied by $|\bm{h}(\omega)|^{2}$: spatial and spectral degrees of freedom are then entangled, and tracing out one leaves the other mixed. Since the collective RX modes are split and broadened on the scale of $\Gamma$ (Section~\ref{sec:results}) and a lifetime-limited photon has the same bandwidth, a photon from an emitter identical to the receiver dots is not narrowband with respect to the collective receiver modes in the geometry studied here. The purity $\Tr\rho_{R}^{2}$ measures the effect. Fig.~\ref{fig:rxstate} shows $\rho_{R}$ for the default link and its purity as a function of the photon linewidth.

\subsection{Spectrally matched receive model}
In the far field $\bm{h}(\omega)\approx\kappa(\omega)\av(\shat)$. The spectral weight $|\kappa(\omega)|^{2}$ follows from $\cv_{T}(\omega)$ and in general contains contributions from several collective TX modes, since the steering state \eqref{eq:TD} is not an eigenvector of $\Mm_{TT}$. When the transmitted state is dominated by a single collective resonance, that weight can be approximated by a Lorentzian with centre $\omega_{c}=\operatorname{Re}(\cv^{H}\Mm_{TT}\cv)$ and FWHM $\gamma=-2\operatorname{Im}(\cv^{H}\Mm_{TT}\cv)=\cv^{H}\bm{\Gamma}_{TT}\cv$, i.e., the collective shift and decay rate of the prepared state. The centre frequency and the width are both proportional to $\Gamma_{T}$, and both can be obtained by spectroscopy of the source. We adopt this single-resonance approximation only to obtain a closed-form manifold; three objects must therefore be kept apart: the exact spectral weight $|\kappa(\omega)|^{2}$ obtained numerically from the TX model, its Lorentzian approximation, and the reduced spectrally matched manifold that follows. Substituting the Lorentzian into \eqref{eq:CRparseval} gives the spectrally matched (SM) model
\begin{equation}
\bm{R}_{\mathrm{SM}}(\shat)=\frac{\int\! S(\omega)\,(\omega\Id-\Mm_{RR})^{-1}\av\av^{H}(\omega\Id-\Mm_{RR})^{-H}d\omega}{\Tr(\cdot)},
\label{eq:RSM}
\end{equation}
with the unit-area Lorentzian $S(\omega)=(\gamma/2\pi)/[(\omega-\omega_{c})^{2}+\gamma^{2}/4]$ and $\av=\av(\shat)$; $\Tr(\cdot)$ denotes the trace of the numerator. The integral converges because the integrand decays as $|\omega|^{-4}$. After the shift $\omega\rightarrow\omega+\omega_{c}$, let $\Mm_{RR}-\omega_{c}\Id=\bm{V}\bm{\Lambda}\bm{V}^{-1}$ with eigenvalues $\lambda_{k}$ and let $\bm{u}=\bm{V}^{-1}\av$. Closing the contour in the upper half plane encloses the poles $\omega=i\gamma/2$ of $S$ and $\omega=\lambda_{l}^{*}$ of the conjugate resolvent (all $\lambda_{l}$ lie in the lower half plane), and the residues give
\begin{gather}
\bm{R}_{\mathrm{SM}}\propto\bm{V}\big[u_{k}u_{l}^{*}\,I_{kl}\big]\bm{V}^{H},\nonumber\\
I_{kl}=\frac{1}{(i\gamma/2-\lambda_{k})(i\gamma/2-\lambda_{l}^{*})}+\frac{i\gamma}{(\lambda_{l}^{*2}+\gamma^{2}/4)(\lambda_{l}^{*}-\lambda_{k})}.
\label{eq:Ikl}
\end{gather}
Evaluating \eqref{eq:Ikl} costs one eigen-decomposition of $\Mm_{RR}$ plus $O(N^{2})$ operations per look direction. For $\gamma\rightarrow0$, $\bm{R}_{\mathrm{SM}}$ collapses to the rank-one narrowband model $\bv\bv^{H}/\|\bv\|^{2}$ with
\begin{equation}
\bv(\shat)=(\omega_{c}\Id-\Mm_{RR})^{-1}\av(\shat),
\label{eq:bca}
\end{equation}
which we call the coupling-aware (CA) manifold; it is the linear-response manifold of a semiclassically driven coupled array, here including the centre-frequency offset $\omega_{c}$. To summarize the status of the three expressions: \eqref{eq:CRparseval}--\eqref{eq:rhoR} define the spectrum-integrated receiver state within the TX--Green-function--RX model and use no spectral or far-field assumption beyond the Markov elimination; \eqref{eq:RSM}--\eqref{eq:Ikl} are a reduced spectrally matched representation obtained after the far-field factorization \eqref{eq:hfar} and the single-resonance spectral approximation; and \eqref{eq:bca} is the further narrowband limit of that reduction. Whether the Lorentzian step is quantitatively harmless for the arrays studied here is checked against the numerically computed spectrum in Section~\ref{sec:results}-A. Pure dephasing of the source broadens the photon to $\gamma+2\gamma_{\phi}$, and \eqref{eq:Ikl} accepts this linewidth directly. The spatially matched (SM) model is a statement about linear electrodynamics and the photon spectrum; a classical receiver facing a broadband classical field could use the same correction.

\section{Readout, Measurement Model and State Reconstruction}\label{sec:meas}
\subsection{Heterodyne snapshots}
Each RX dot is coupled, with equal efficiency and negligible crosstalk, to its own single-mode collection channel (a local waveguide or grating coupler), so the channel amplitude is proportional to $\hat{\sigma}^{-}_{n}$; the $N$ outputs are mixed with a common local oscillator on balanced heterodyne detectors and integrated over a window long compared with $T_{1}$, which traces out time as in \eqref{eq:rhoR}. 
One repetition of the link produces one snapshot $\bm{x}[k]\in\mathbb{C}^{N}$, $k=1,\dots,K$, whose element $x_{n}[k]$ is the complex heterodyne output of channel $n$, scaled so that vacuum gives $E|x_{n}|^{2}=1$; with probability $P_{d}$ it contains one photon in the spatial state $\rho_{R}$, otherwise only vacuum.\\
It is to note here that $P_{\mathrm{abs}}$ and $P_{d}$ are different quantities. $P_{\mathrm{abs}}$ is the absorption probability of the bare link, $1.0\times10^{-4}$ at $R=200\lambda_{0}$ (Section~\ref{sec:results}). $P_{d}$ is the probability that a \emph{retained} snapshot contains a detected photon, that is, a probability conditioned on the gating or heralding rule that selects which repetitions are kept. Writing $P_{\mathrm{g}}$ for the fraction of repetitions retained and $\eta_{\mathrm{c}},\eta_{\mathrm{d}}$ for the collection and detection efficiencies, $P_{d}=\min\{1,\ P_{\mathrm{abs}}\eta_{\mathrm{c}}\eta_{\mathrm{d}}/P_{\mathrm{g}}\}$, which satisfies $0\leq P_{d}\leq1$ because the retained fraction cannot be smaller than the fraction of repetitions that actually produce a detection. The two limits used below are therefore $P_{d}\ll1$, a loosely gated receiver in which most snapshots are vacuum, and $P_{d}=1$, a perfectly heralded receiver in which every retained snapshot carries a photon; neither is the bare-link absorption probability. The Monte Carlo results are conditioned on the retained gated or heralded snapshot ensemble rather than on the total number of transmitted photons.\\
Heterodyne outcomes are distributed according to the Husimi $Q$ function of the detected field \cite{shapiro1984,leonhardt1997}. For the mixture of vacuum and a one-photon state with spatial density matrix $\rho_{R}$ the density of one snapshot is
\begin{equation}
p(\bm{x};\shat)=\pi^{-N}e^{-\|\bm{x}\|^{2}}\big[(1-P_{d})+P_{d}\,\bm{x}^{H}\rho_{R}(\shat)\bm{x}\big],
\label{eq:pdf}
\end{equation}
with covariance and sample estimate
\begin{equation}
\bm{R}_{x}=E[\bm{x}\bm{x}^{H}]=\Id+P_{d}\,\rho_{R},~ \hat{\bm{R}}_{x}=\frac{1}{K}\sum_{k=1}^{K}\bm{x}[k]\bm{x}^{H}[k].
\label{eq:Rx}
\end{equation}
The identity in $\bm{R}_{x}$ is the vacuum (shot) noise that heterodyne detection adds to every channel, one unit of noise per mode; it plays the part of receiver noise. Detector excess noise of variance $\sigma_{e}^{2}$ would add $\sigma_{e}^{2}\Id$. The density \eqref{eq:pdf} is not necessarily Gaussian, and this matters for the bound in Section~\ref{sec:meas}-C. Equation \eqref{eq:Rx} follows from $E[\bm{w}\bm{w}^{H}|\bm{u}^{H}\bm{w}|^{2}]=\Id+\bm{u}\bm{u}^{H}$ for $\bm{w}\sim\mathcal{CN}(\bm{0},\Id)$: the photon adds one unit of variance in its own mode. Snapshots are drawn exactly from \eqref{eq:pdf}, a vacuum snapshot being $\bm{w}\sim\mathcal{CN}(\bm{0},\Id)$ and a photon snapshot $\bm{x}=\bm{w}+(z-\bm{u}^{H}\bm{w})\bm{u}$, where the mode $\bm{u}$ is drawn from the eigen-decomposition of $\rho_{R}$ with probabilities equal to its eigenvalues, $|z|^{2}\sim\mathrm{Gamma}(2,1)$ and $\arg z$ is uniform.

\subsection{State reconstruction}
The DOA branch returns a direction; the second branch of Fig.~\ref{fig:arch} returns an estimate $\hat{\rho}_{R}$ of the true state $\rho_{R}$. From \eqref{eq:Rx}, $\hat{\bm{R}}_{x}-\Id$ is an unbiased estimate of $P_{d}\rho_{R}$. Its trace estimates $P_{d}$, so dividing by the trace removes the unknown $P_{d}$. Sampling noise leaves the normalized matrix Hermitian but generally with some negative eigenvalues. We keep its eigenvectors and replace its eigenvalues by their Euclidean projection onto the probability simplex (non-negative, summing to one). For additive Gaussian noise this projection is the maximum-likelihood physical state \cite{smolin2012}. The quality of $\hat{\rho}_{R}$ is measured by the Uhlmann fidelity \cite{jozsa1994}
\begin{equation}
F(\rho_{R},\hat{\rho}_{R})=\Big(\Tr\sqrt{\sqrt{\rho_{R}}\,\hat{\rho}_{R}\sqrt{\rho_{R}}}\Big)^{2},
\label{eq:fid}
\end{equation}
in the squared convention, so that $0\leq F\leq1$ and $F=1$ only for $\hat{\rho}_{R}=\rho_{R}$. The same definition is used for every fidelity in the paper, including the fidelity of $\rho_{R}$ to the FS, CA and SM models.

\subsection{Exact and Gaussian Cram\'er--Rao bounds}
For $K$ independent snapshots and the parameter vector $\bm{\vartheta}=(\theta,\phi)$ with $\partial_{i}=\partial/\partial\vartheta_{i}$, the Fisher information matrix of the snapshot density \eqref{eq:pdf}, exact for that density and not a quantum Fisher information over all possible measurements, is
\begin{equation}
[\bm{F}]_{ij}=K\,E\!\left[\frac{P_{d}^{2}\,(\bm{x}^{H}\partial_{i}\rho_{R}\bm{x})(\bm{x}^{H}\partial_{j}\rho_{R}\bm{x})}{\big[(1-P_{d})+P_{d}\,\bm{x}^{H}\rho_{R}\bm{x}\big]^{2}}\right],
\label{eq:FIMexact}
\end{equation}
which we evaluate by Monte Carlo with $2\times10^{5}$ samples and central-difference derivatives of $\rho_{R}(\shat)$. For comparison, a zero-mean complex Gaussian vector with the same covariance $\bm{R}_{x}$ has $[\bm{F}_{G}]_{ij}=K\Tr(\bm{R}_{x}^{-1}\partial_{i}\bm{R}_{x}\bm{R}_{x}^{-1}\partial_{j}\bm{R}_{x})$. We call $\bm{F}^{-1}$ the \emph{exact CRB} and $\bm{F}_{G}^{-1}$ the \emph{Gaussian CRB}. Under the regularity conditions of the Fisher-information formulation, the exact CRB bounds unbiased estimators based on the full snapshot distribution \eqref{eq:pdf}; the Gaussian CRB is the corresponding asymptotic bound for estimators that use only $\hat{\bm{R}}_{x}$ and treat the data as Gaussian. Both are classical Fisher-information bounds for the specified heterodyne measurement and are not quantum Fisher information bounds optimized over all measurements. Both are reported as great-circle standard deviations, $\mathrm{CRB}=\sqrt{[\bm{F}^{-1}]_{\theta\theta}+\sin^{2}\theta\,[\bm{F}^{-1}]_{\phi\phi}}$, so a ratio of two such values is a ratio of standard deviations and the variance ratio is its square.

\section{Direction Finding With the Coupled Quantum-Dot Receiver}\label{sec:doa}
The estimators below are not generic array processing applied to an unrelated data model. Every quantity they use comes from Sections~\ref{sec:rx} and \ref{sec:meas}: the data covariance is $\bm{R}_{x}=\Id+P_{d}\rho_{R}(\shat)$ of \eqref{eq:Rx}, in which the signal subspace is the \emph{receiver state} \eqref{eq:rhoR} produced by the Green-function link, and the candidate manifolds are the directional models of that same state. The DOA problem is therefore one of matching the modeled $\rho_{R}(\shat)$ to the measured covariance, and a manifold is wrong exactly when the photon it was derived for differs from the photon that arrives.

\subsection{Manifolds}
Three receive manifolds are compared, each normalized to unit norm.
\emph{FS} is the free-space plane-wave vector $\av(\shat)$ of \eqref{eq:a}, which ignores $\bm{G}_{RR}$; \emph{CA} is the narrowband vector $\bv(\shat)$ of \eqref{eq:bca}, exact for a monochromatic field by Proposition~\ref{prop:1}; \emph{SM}: the matrix $\bm{R}_{\mathrm{SM}}(\shat)$ of \eqref{eq:Ikl} is the full spectrum-weighted spatial state predicted for direction $\shat$, and for the subspace estimators below, which require a vector-valued manifold, we use its principal eigenvector $\bm{e}_{1}(\shat)$; this is the step that adds the centre frequency and linewidth of the photon to the manifold. All three are tabulated on a grid; CA and SM need $\Mm_{RR}$, from the geometry through \eqref{eq:JG} or from calibration. FS and CA are the manifolds a classical receiver would use with and without mutual-coupling compensation \cite{friedlander1991}; SM is the manifold that the single-photon model of Section~\ref{sec:rx} actually predicts.

\subsection{Estimators}
\begin{figure}[!t]
\centering
\includegraphics[width=\columnwidth]{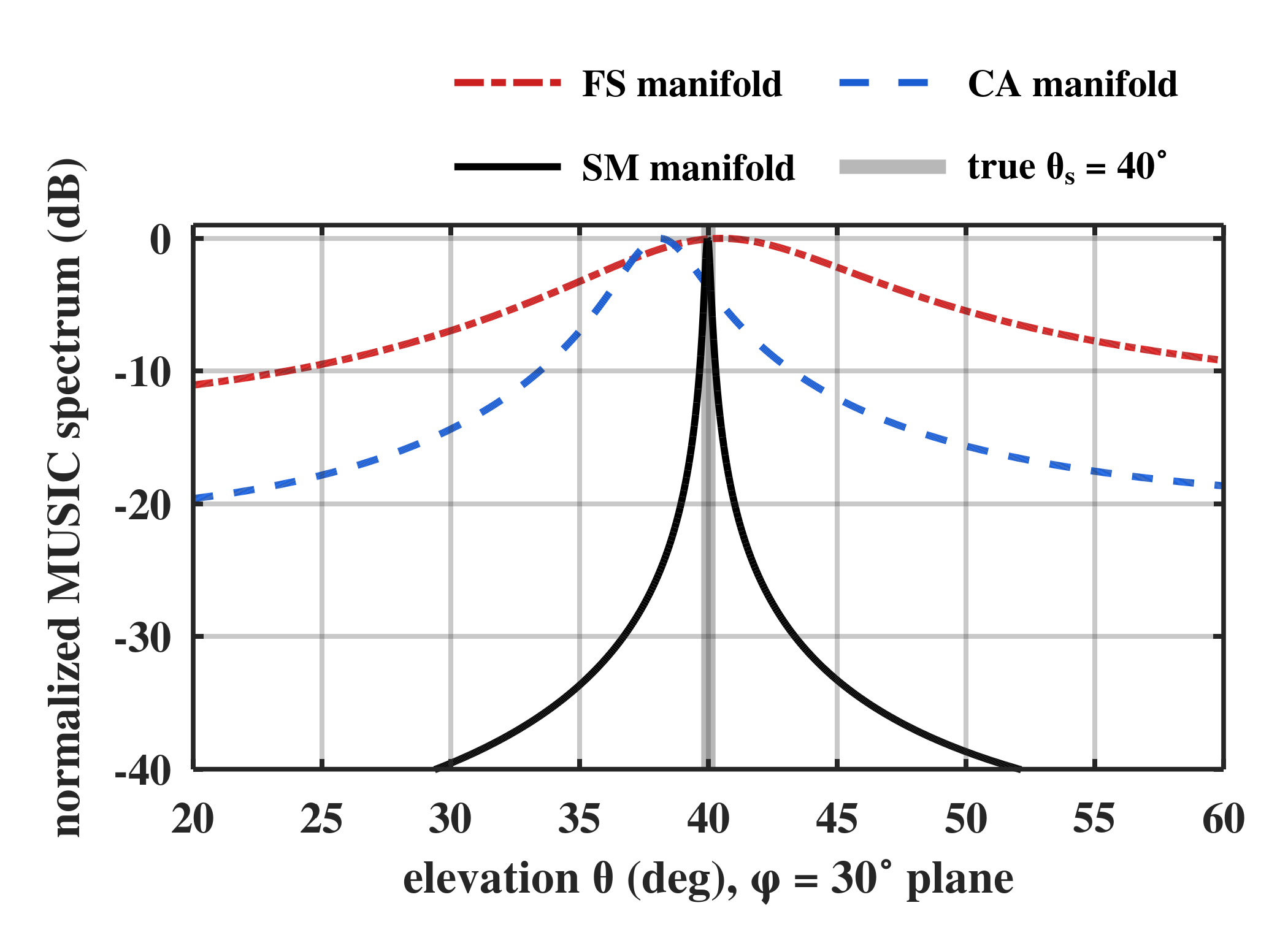}
\caption{Normalized MUSIC spectra for the exact link covariance ($P_{d}=0.1$) in the $\phi=30^{\circ}$ plane with the free-space (FS), coupling-aware (CA) and spectrally matched (SM) manifolds; the grey vertical line marks the true $\theta_{s}=40^{\circ}$. Bartlett and MVDR spectra (not shown) vary by less than \SI{0.4}{\decibel} over this cut at $P_{d}=0.1$; their errors are given in Section~\ref{sec:results}-F.}
\label{fig:spectra}
\end{figure}

With $\bm{g}(\shat)$ denoting any of the three manifolds, we use the Bartlett spectrum $P_{\mathrm{B}}=\bm{g}^{H}\hat{\bm{R}}_{x}\bm{g}$, the MVDR  spectrum $P_{\mathrm{MV}}=(\bm{g}^{H}\hat{\bm{R}}_{x}^{-1}\bm{g})^{-1}$ \cite{capon1969} and the MUSIC spectrum $P_{\mathrm{MU}}=(\bm{g}^{H}\hat{\bm{E}}_{n}\hat{\bm{E}}_{n}^{H}\bm{g})^{-1}$ \cite{schmidt1986}. Here $\hat{\bm{E}}_{n}$ holds the $N-Q$ eigenvectors of $\hat{\bm{R}}_{x}$ with the smallest eigenvalues, and $Q$ is the number of sources. The spectra are evaluated on a $2^{\circ}\times2^{\circ}$ grid and the $Q$ largest well-separated maxima refined by a Nelder--Mead search of the continuous spectrum. The search is restricted to the upper half space because of the mirror symmetry \eqref{eq:mirror}: a planar array of in-plane dipoles cannot tell $\theta$ from $180^{\circ}-\theta$. Fig.~\ref{fig:spectra} shows MUSIC spectra with each of the three manifolds.

\subsection{Shift invariance and ESPRIT}
ESPRIT \cite{roy1989} requires the manifold to be invariant under a translation of a subarray. The FS vector of a uniform planar array has this property. The CA vector $\Mm_{R}^{-1}\av$ does not, because edge and interior dots sit in different coupling environments, so ESPRIT applied directly to coupled data is biased. For a narrowband field the invariance is restored exactly by \emph{decoupling}: the signal eigenvector is multiplied by $(\Mm_{RR}-\omega_{c}\Id)$ before the least-squares shift equations in $x$ and $y$ are formed \cite{friedlander1991}. For a photon of finite bandwidth the signal subspace is not spanned by a single CA vector, so decoupling is exact only as $\gamma\rightarrow0$. The decoupled 2-D ESPRIT is included to measure that residual. Beamspace processing is not pursued, since with $N=16$ the element-space eigen-decomposition is already cheap.

\subsection{Receiver procedure and complexity}
The SM receiver needs three one-time steps -- diagonalizing $\Mm_{RR}$ at cost $O(N^{3})$, measuring $\omega_{c}$ and $\gamma$ once per source type, and storing the principal eigenvector of $\bm{R}_{\mathrm{SM}}$ on the search grid -- and one per-data step, forming $\hat{\bm{R}}_{x}$ at cost $O(KN^{2})$ and searching the stored table. Only the last depends on the data. Once the spectrally matched manifold has been precomputed and tabulated, the per-snapshot computational structure is therefore comparable to conventional subspace DOA estimation with a stored calibration table \cite{friedlander1991,vantrees2002}; the total cost is larger by the offline eigen-decomposition and the grid tabulation, which are performed once per array and source type.
\section{Numerical Results}\label{sec:results}
All results use the parameters of Table~\ref{tab:param} and come from the accompanying MATLAB/Octave package. Angular errors are great-circle angles. Unless stated otherwise, link DOA errors are computed from the exact covariance $\bm{R}_{x}=\Id+P_{d}\rho_{R}$ at $P_{d}=0.1$, i.e., for $K\rightarrow\infty$; they are pure model-mismatch biases. FS, CA and SM keep the same colours and line styles in every figure. Sections~\ref{sec:results}-C report transmitter quantities that are classically equivalent and serve as validation; Sections~\ref{sec:results}-E--H report the receive behaviour that the classical model does not reproduce.

\subsection{Numerical verification}
Relative agreement is $2.3\times10^{-15}$ between \eqref{eq:X} and the Kronecker-form Liouvillian solution, $2.1\times10^{-12}$ between the dephasing solver at $\gamma_{\phi}\rightarrow0$ and the undephased result, and $1.8\times10^{-10}$ between \eqref{eq:Ikl} and quadrature of \eqref{eq:RSM}; the radiated photon number equals one to within $10^{-6}$ and the operator-algebra $g^{(2)}$ reproduces \eqref{eq:g2}. The last of these figures verifies the closed-form integration of \eqref{eq:Ikl}; the Lorentzian approximation itself is checked separately. Computing the spectral weight directly from the TX model, $|\bm{h}(\omega)|^{2}$ with $\bm{h}(\omega)=\Mm_{RT}\cv_{T}(\omega)$, and normalizing it to unit area gives a line of FWHM $0.975\,\Gamma$ against $\gamma=0.891\,\Gamma$ for the single-resonance Lorentzian built from the same state; the two normalized weights differ by $5.4\%$ in normalized RMS error and overlap by $(\int\sqrt{S_{\mathrm{num}}(\omega)S(\omega)}\,d\omega)^{2}=0.9994$. Replacing the Lorentzian in \eqref{eq:RSM} by the numerical weight and repeating the default-link MUSIC estimate of Section~\ref{sec:results}-F gives $0.015^{\circ}$, against $0.031^{\circ}$ for the closed-form SM manifold and $1.813^{\circ}$ for CA; the quadrature is converged to $\pm0.001^{\circ}$ for grids of $1200$ points or more over $\pm60\,\Gamma$. Roughly half of the small SM residual is therefore attributable to the single-resonance approximation, and none of the two orders of magnitude that separate SM from CA. The two analytic limits of the model, an exact CA manifold for $\Gamma_{T}\rightarrow0$ and identical manifolds for an uncoupled receiver, are recovered as well (Section~\ref{sec:results}-F).

\subsection{Mutual coupling and manifold distortion}
Fig.~\ref{fig:coupling}(a) shows the pair couplings \eqref{eq:JG} for separations along and across the dipole axis; at $d=\lambda_{0}/2$ ($k_{0}r=\pi$) both $|\Gamma_{mn}|/\Gamma$ and $|J_{mn}|/\Gamma$ are a sizeable fraction of one, the collinear pair being dominated by near-field terms. The collective decay rates [Fig.~\ref{fig:coupling}(b)] span $0.154\Gamma$ to $2.426\Gamma$ at $d=\lambda_{0}/2$ and the shifts $-0.611\Gamma$ to $0.371\Gamma$. In DOA terms [Fig.~\ref{fig:coupling}(c)], the CA--FS mismatch $\epsilon_{m}=1-|\av^{H}\bv|^{2}/(\|\av\|^{2}\|\bv\|^{2})$ is $0.109$ at $d=\lambda_{0}/2$, where FS-MUSIC is biased by $3.09^{\circ}$; the bias peaks at $16.2^{\circ}$ near $0.30\lambda_{0}$ ($\epsilon_{m}=0.668$) and falls below $1^{\circ}$ only for $d\geq0.57\lambda_{0}$. Coupling cannot be ignored at half-wavelength spacing.

\subsection{Transmitter: validation, steering and photon statistics}
A four-element array along $z$ with $\hat{\bm{x}}$ dipoles reproduces the classical equivalence of Section~\ref{sec:tx} (Fig.~\ref{fig:linear}): the $W$ state ($c_{n}=1/\sqrt{N}$) and the uniformly fed classical array have identical patterns, with $D_{0}=8.74$~dBi, half-power beamwidth (HPBW) $26.32^{\circ}$ and side-lobe level (SLL) $-11.30$~dB, while the state steered to endfire at $d=0.3\lambda_{0}$ gives $D_{0}=7.75$~dBi, HPBW $103.3^{\circ}$, front-to-back ratio $12.04$~dB and pattern asymmetry $\eta_{A}=\int|G^{(1)}(\theta)-G^{(1)}(\pi-\theta)|d\theta/\int G^{(1)}d\theta=1.806$; at $d=0.5\lambda_{0}$ the same phase law gives the familiar symmetric bidirectional pattern. None of this is quantum.\\ To demonstrate the planar steering, Fig.~\ref{fig:patterns} shows prepared-state patterns on the $(u,v)$ disc; Fig.~\ref{fig:scan} and Table~\ref{tab:steer} summarize the scan at $\phi_{0}=45^{\circ}$. Broadside gives $D_{0}=13.92$~dBi and HPBW $25.4^{\circ}$; scanning to $\theta_{0}=60^{\circ}$ lowers $D_{0}$ to $11.79$~dBi, widens the beam to $52.9^{\circ}$ and grows $\Delta\Omega$ from $0.27^{\circ}$ at $5^{\circ}$ to $5.34^{\circ}$ [Fig.~\ref{fig:scan}(b),(c)], the peak moving toward broadside and toward $\phi=90^{\circ}$ where the $\hat{\bm{x}}$ dipole radiates most strongly, as Section~\ref{sec:tx} predicts. The emitted-photon directivity stays within $0.1$~dB of $D_{0}$ up to $45^{\circ}$ and $0.39$~dB behind at $60^{\circ}$, and the steering states, not being eigenmodes, decay at $0.93\Gamma$ to $1.59\Gamma$.

\begin{table}[ht!]
\caption{Steering of the $4\times4$ Array: Prepared State ($D_{0}$) and Emitted Photon ($D_{0}^{\mathrm{int}}$)}
\label{tab:steer}
\centering
\footnotesize
\setlength{\tabcolsep}{2.2pt}
\begin{tabular}{@{}lccccccc@{}}
\toprule
$(\theta_{0},\phi_{0})$ & peak & $\Delta\Omega$ & $D_{0}$ & $D_{0}^{\mathrm{int}}$ & HPBW & SLL & $\Gamma_{\mathrm{st}}/\Gamma$\\
 & (deg) & (deg) & (dBi) & (dBi) & (deg) & (dB) & \\
\midrule
$(0,0)$ & $(0.00,0.00)$ & 0.00 & 13.92 & 13.91 & 25.4 & $-14.40$ & 0.974\\
$(20,0)$ & $(18.29,0.00)$ & 1.71 & 13.63 & 13.61 & 26.5 & $-11.50$ & 0.929\\
$(30,45)$ & $(28.67,47.48)$ & 1.80 & 13.26 & 13.18 & 30.3 & $-19.41$ & 1.003\\
$(45,90)$ & $(45.00,90.00)$ & 0.00 & 11.80 & 11.88 & 40.5 & $-5.31$ & 1.587\\
$(60,135)$ & $(55.35,131.89)$ & 5.34 & 11.79 & 11.40 & 52.9 & $-1.91$ & 1.057\\
$(50,315)$ & $(46.91,312.12)$ & 3.76 & 12.04 & 11.87 & 42.0 & $-4.45$ & 1.104\\
\bottomrule
\end{tabular}
\end{table}

Table~\ref{tab:g2} gives $g^{(2)}$ in the commanded direction. A pure single-excitation state is perfectly antibunched, and a $10\%$ two-excitation admixture raises $g^{(2)}$ only to $0.136$ ($N=4$) and $0.159$ ($N=16$), far below the coherent-state reference. This is the principal nonclassical distinction considered for the transmitter in this model, and the results below show that direction finding does not use it. Under pure dephasing of every dot (Fig.~\ref{fig:txdeph}) TX, $D_{0}^{\mathrm{int}}$ of the $(30^{\circ},45^{\circ})$ state falls by $1.13$~dB (from $13.17$ to $12.04$~dBi) at the resonant bound $\gamma_{\phi}=0.167\Gamma$ and by $6.20$~dB (to $6.97$~dBi) at the non-resonant $2.24\Gamma$, while the coherence ratio $\eta_{\mathrm{coh}}$ -- the $\ell_{1}$ norm of the off-diagonal part of the normalized $\bm{X}$, relative to its undephased value -- drops to $0.715$ and $0.181$. The beam direction hardly moves ($0.2^{\circ}$ over the whole sweep) while the HPBW grows from $30.2^{\circ}$ to $36.0^{\circ}$: dephasing costs directional contrast, not pointing, because it suppresses the off-diagonal entries of $\RQ$ without biasing their phases. Reading the sweep as a specification, a $0.5$~dB directivity budget is met at $\gamma_{\phi}=0.05\Gamma$ ($0.38$~dB, $\eta_{\mathrm{coh}}=0.884$) but not at $0.1\Gamma$ ($0.72$~dB, $\eta_{\mathrm{coh}}=0.801$), so the transmitter must be driven resonantly and kept near the transform limit; this is a budget derived from a chosen criterion, not a physical boundary.

\subsection{TX-RX communication channel}
Fig.~\ref{fig:link} varies the communication link distance from $3\lambda_{0}$ to $1000\lambda_{0}$; this studies the channel and the manifold, not a practical long-distance 1550-nm link. The plane-wave error \eqref{eq:epsch} follows the $R^{-2}$ reference line from $1.5\times10^{-1}$ to $1.8\times10^{-6}$ ($4.5\times10^{-5}$ at $R=200\lambda_{0}$), and the absorption probability the same law, from $0.22$ to $4.0\times10^{-6}$ ($1.01\times10^{-4}$ at $200\lambda_{0}$). The DOA errors [Fig.~\ref{fig:link}(b)] show two regimes. For $R\leq10\lambda_{0}$ the incident field is not a plane wave and no plane-wave-based manifold is adequate (SM-MUSIC errs by $3.72^{\circ}$ at $3\lambda_{0}$). Beyond $20\lambda_{0}$ the RX coupling alone sets the error: FS levels off near $1.67^{\circ}$ and CA near $1.81^{\circ}$, while SM stays between $0.014^{\circ}$ and $0.042^{\circ}$, about half of that residual coming from the single-Lorentzian description of the TX spectrum (Section~\ref{sec:results}-A). The RX purity is $0.949$ for every $R\geq20\lambda_{0}$, so the mixedness is set by the RX coupling and the photon spectrum, not the channel.

\subsection{Receiver state}
Fig.~\ref{fig:rxstate} shows the receiver state $\rho_{R}$ of \eqref{eq:rhoR} for the default link. The magnitudes $|[\rho_{R}]_{mn}|$ are dominated by the plane-wave phase gradient, modulated at the edges by the coupling. The eigenvalues of $\rho_{R}$ are $0.974$, $0.022$ and $0.004$, followed by smaller ones, and its purity is $0.949$. The fidelity \eqref{eq:fid} of $\rho_{R}$ is $0.918$ to the FS model $\av\av^{H}/N$, $0.955$ to the narrowband CA model, and $0.99997$ to the SM model of \eqref{eq:Ikl}. Fig.~\ref{fig:rxstate}(c) contains the central result of the paper. The radiative rate $\Gamma_{T}$ of the TX dots, and with it the photon linewidth, is varied while the RX is left unchanged; $\Gamma_{T}=\Gamma$ is the lifetime-limited photon of an identical emitter and $\Gamma_{T}\rightarrow0$ the narrowband limit of Proposition~\ref{prop:1}. As $\Gamma_{T}/\Gamma$ rises from $0.01$ to $2$ the purity falls from $0.998$ to $0.943$ and the CA error grows from $0.038^{\circ}$ to $2.05^{\circ}$, while the FS error falls from $3.05^{\circ}$ to $1.43^{\circ}$ because averaging over a wider band washes out the frequency-selective coupling; the curves cross near $\Gamma_{T}\approx0.8\Gamma$, leaving the lifetime-limited photon on the wrong side for CA, and SM stays below $0.036^{\circ}$ throughout. Every photon in this sweep is a single photon, so photon statistics cannot cause the effect: the narrowband manifold fails because of the photon bandwidth.

\subsection{DOA spectra and the four-system baseline}
Fig.~\ref{fig:spectra} shows MUSIC spectra in the $\phi=30^{\circ}$ plane and Table~\ref{tab:baseline} the errors of all estimators. For a given manifold the three spectral estimators agree -- FS $1.64^{\circ}$--$1.67^{\circ}$, CA $1.80^{\circ}$--$1.81^{\circ}$, SM $0.031^{\circ}$ -- so the manifold matters far more than the estimator. The Bartlett and MVDR spectra vary by only a few tenths of a decibel over the cut, since at $P_{d}=0.1$ the signal term is one tenth of the vacuum noise, yet their peaks are placed correctly with the SM manifold. Direct ESPRIT errs by $1.95^{\circ}$ and decoupled ESPRIT by $0.58^{\circ}$, the residual left by the finite photon bandwidth in the shift-invariance structure (Section~\ref{sec:doa}-C).\\
The four-system baseline in the lower half of Table~\ref{tab:baseline} separates the ingredients, the narrowband system using $\Gamma_{T}=10^{-3}\Gamma$, which by Proposition~\ref{prop:1} responds as to a weak monochromatic drive. With an uncoupled RX all manifolds are exact to within the $0.009^{\circ}$ residual of the finite link distance; with a coupled RX, FS is biased by $3.09^{\circ}$ while CA is exact ($0.005^{\circ}$); a single photon on an uncoupled RX is again unbiased ($0.014^{\circ}$). Only a broadband single photon on a coupled RX breaks the CA manifold: collective electrodynamics acting on a finite-bandwidth wavepacket, not photon statistics.

\begin{table}[!t]
\caption{DOA Error (deg) for the Exact Link Covariance, Source $(40^{\circ},30^{\circ})$, $R=200\lambda_{0}$. FS: Free-Space; CA: Coupling-Aware; SM: Spectrally Matched. Dashes Denote Combinations Not Evaluated}
\label{tab:baseline}
\centering
\footnotesize
\setlength{\tabcolsep}{3pt}
\begin{tabular}{@{}lcccc@{}}
\toprule
Estimator & FS & CA & SM & \\
\midrule
Bartlett & 1.639 & 1.799 & 0.031 & \\
MVDR & 1.636 & 1.798 & 0.031 & \\
MUSIC & 1.668 & 1.813 & 0.031 & \\
ESPRIT, direct$^{\dagger}$ & 1.948 & -- & -- & \\
ESPRIT, decoupled$^{\ddagger}$ & -- & 0.577 & -- & \\
\midrule
TX / RX (MUSIC) & FS & CA & SM & purity\\
\midrule
Narrowband ($\Gamma_{T}=10^{-3}\Gamma$) / uncoupled & 0.009 & 0.009 & 0.009 & 1.000\\
Narrowband ($\Gamma_{T}=10^{-3}\Gamma$) / coupled & 3.085 & 0.005 & 0.009 & 1.000\\
Single photon ($\Gamma_{T}=\Gamma$) / uncoupled & 0.014 & 0.014 & 0.014 & 1.000\\
Single photon ($\Gamma_{T}=\Gamma$) / coupled & 1.667 & 1.813 & 0.031 & 0.949\\
\bottomrule
\multicolumn{5}{@{}p{78mm}@{}}{\footnotesize $^{\dagger}$uses the shift invariance of the FS manifold; $^{\ddagger}$decoupled with $\Mm_{RR}-\omega_{c}\Id$, i.e.\ the CA model.}\\
\end{tabular}
\end{table}

\begin{figure*}[!t]
\centering
\includegraphics[width=0.76\textwidth]{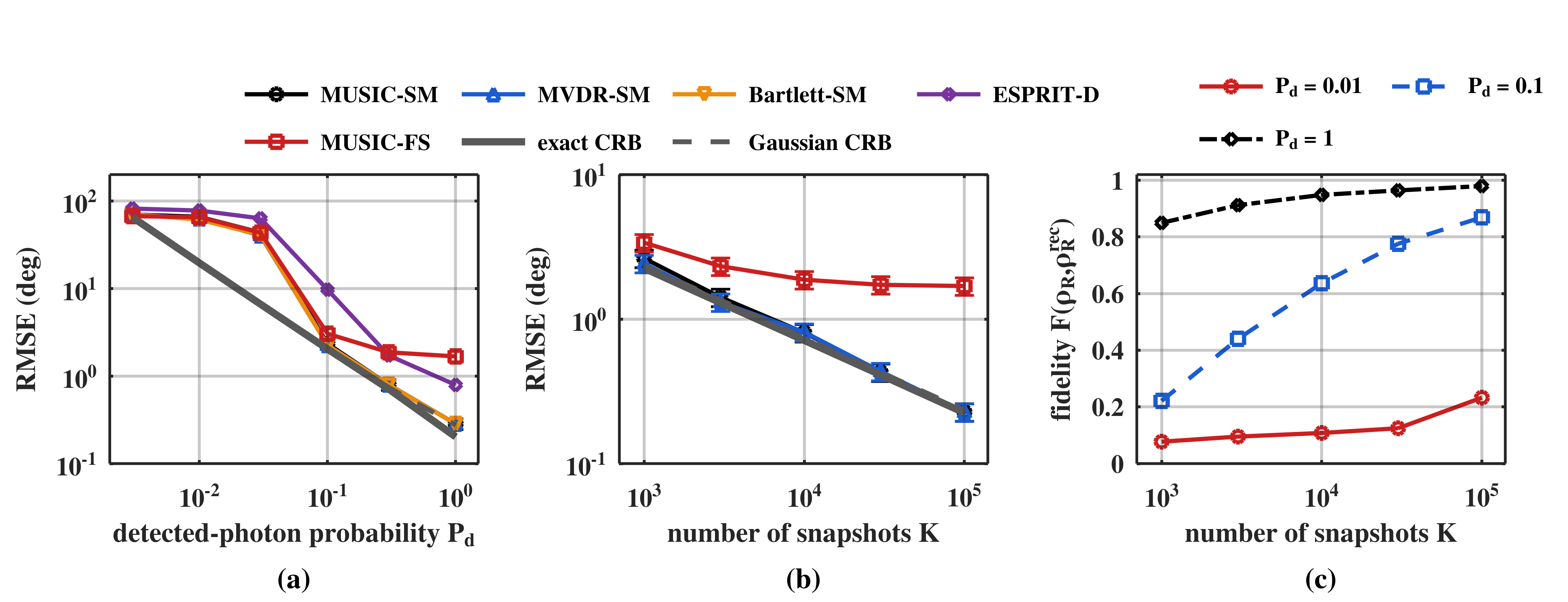}
\caption{Monte Carlo performance with heterodyne snapshots (100 trials per point; $P_{d}$ is the detected-photon probability per gated or heralded snapshot). SM, FS: spectrally matched and free-space manifolds; ESPRIT-D: decoupled ESPRIT. (a) Great-circle RMSE versus $P_{d}$ at $K=10^{4}$, with the exact (thick solid) and Gaussian (dashed) CRB. (b) RMSE versus $K$ at $P_{d}=0.3$; error bars: approximate $95\%$ confidence interval. (c) Fidelity of $\hat{\rho}_{R}$ to $\rho_{R}$ (50 trials).}
\label{fig:rmse}
\end{figure*}

\begin{figure}[!t]
\centering
\includegraphics[width=\columnwidth]{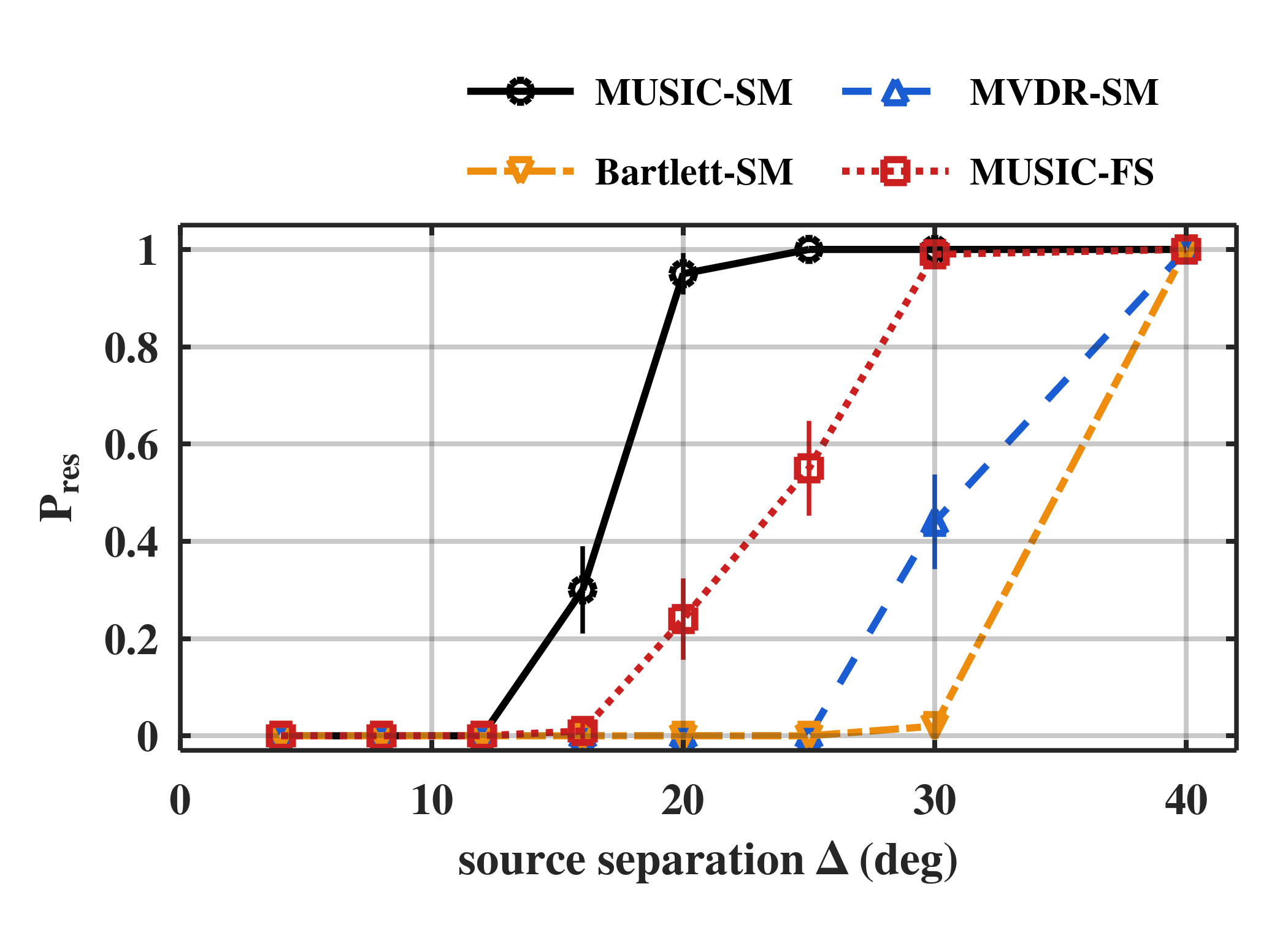}
\caption{Probability of resolving two incoherent single-photon sources at $(40^{\circ}\mp\Delta\theta/2,30^{\circ})$; $K=10^{4}$, $P_{d}=1$, 100 trials per point, $95\%$ binomial error bars. Resolved: both estimates within $\Delta\theta/2$ of distinct true directions. Specific to this setup, not a universal limit.}
\label{fig:resolution}
\end{figure}

\begin{figure*}[!t]
\centering
\includegraphics[width=0.76\textwidth]{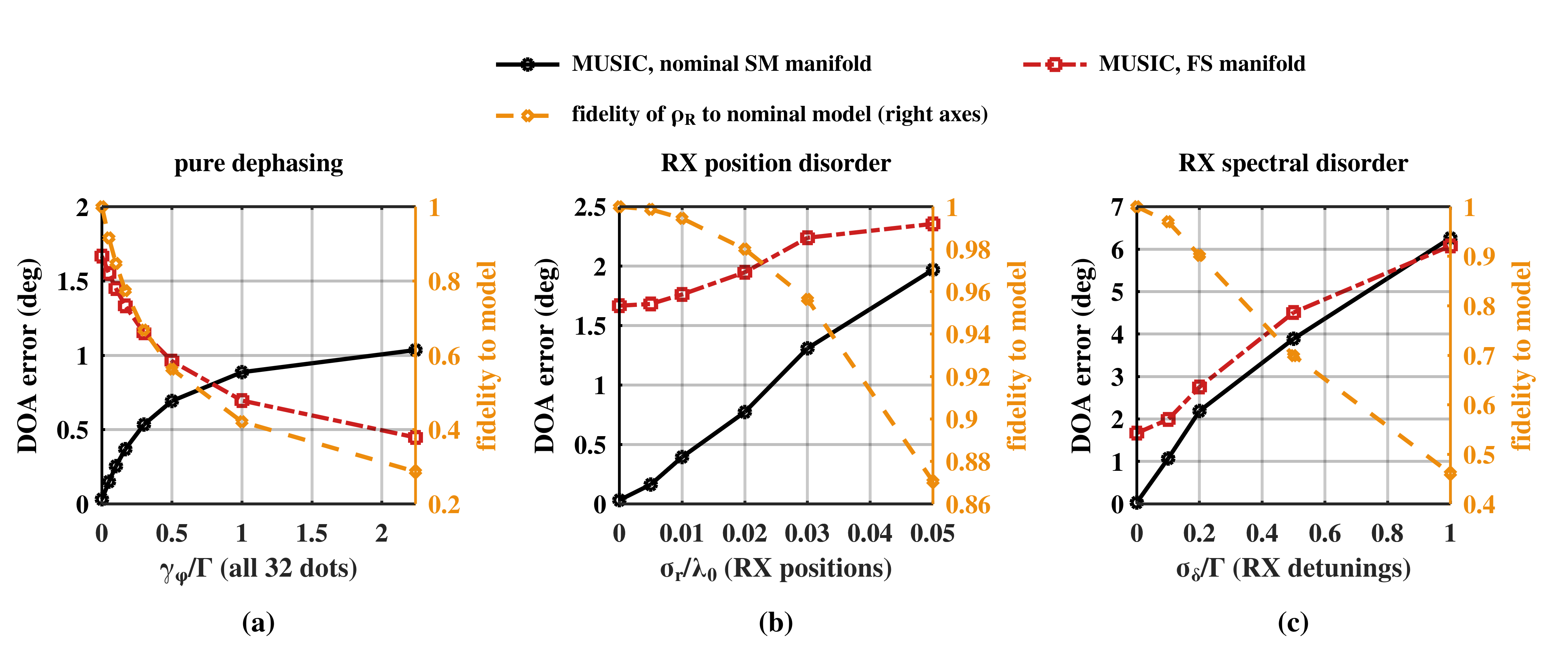}
\caption{Robustness when the \emph{nominal} (ideal-array) SM manifold is used under mismatch ($P_{d}=0.1$): MUSIC error with the nominal spectrally matched (SM) and free-space (FS) manifolds (left axes) and fidelity of $\rho_{R}$ to the nominal model (right axes, dashed). (a) Pure dephasing of all 32 dots, the SM model using the broadened linewidth $\gamma+2\gamma_{\phi}$. (b) RX position disorder $\sigma_{r}$ per coordinate. (c) RX spectral disorder $\sigma_{\delta}$; in (b), (c) errors are RMS and fidelities means over 30 realizations.}
\label{fig:robust}
\end{figure*}

\subsection{Statistical performance, Cram\'er--Rao bounds and reconstruction}
Fig.~\ref{fig:rmse} gives Monte Carlo results from 100 independent sets of $K$ snapshots drawn from \eqref{eq:pdf}, conditioned on gated or heralded repetitions (Section~\ref{sec:meas}); the error bars are the approximate $95\%$ confidence interval of an RMSE from 100 trials, $\pm1.96/\sqrt{200}\approx\pm14\%$. With $K=10^{4}$ [Fig.~\ref{fig:rmse}(a)] all estimators show the usual threshold: below $P_{d}\approx0.05$ the signal term $P_{d}\rho_{R}$ is buried in the vacuum noise and the CRB is not informative. Above it the three SM estimators track the bounds, with RMSEs of $2.25^{\circ}$--$2.34^{\circ}$ against an exact CRB of $2.04^{\circ}$ at $P_{d}=0.1$ and $0.80^{\circ}$ against $0.71^{\circ}$ at $P_{d}=0.3$, while FS-MUSIC levels off at its $1.67^{\circ}$ bias and decoupled ESPRIT approaches its $0.58^{\circ}$ residual ($0.79^{\circ}$ at $P_{d}=1$).\\
At $P_{d}=1$ the SM estimators reach $0.284^{\circ}$, almost on the Gaussian CRB ($0.276^{\circ}$) but $41\%$ above the exact CRB ($0.202^{\circ}$), which is $27\%$ lower in standard deviation, a factor $1.86$ ($2.7$~dB) in variance. The heterodyne data of a nearly pure one-photon state are strongly non-Gaussian, so part of their Fisher information is not carried by the covariance and only a maximum-likelihood estimator built on \eqref{eq:pdf} could recover this margin. At $P_{d}=0.3$ [Fig.~\ref{fig:rmse}(b)] the SM-MUSIC error decreases as $K^{-1/2}$, from $2.63^{\circ}$ at $K=10^{3}$ to $0.228^{\circ}$ at $K=10^{5}$, which is within $1.5\%$ of the exact CRB ($0.224^{\circ}$); FS-MUSIC stalls at $1.69^{\circ}$. Reconstructing the state [Fig.~\ref{fig:rmse}(c)] is far harder than finding the direction, since all $N^{2}-1=255$ real parameters of $\rho_{R}$ must be resolved rather than two angles: the fidelity rises from $0.850$ to $0.980$ between $K=10^{3}$ and $10^{5}$ at $P_{d}=1$, reaches $0.869$ at $P_{d}=0.1$, and stays below $0.24$ at $P_{d}=0.01$. The DOA branch needs one to two orders of magnitude fewer detected photons than the state branch.

\subsection{Two-source resolution}
Fig.~\ref{fig:resolution} gives the probability of resolving two mutually incoherent single-photon sources at $(40^{\circ}\mp\Delta\theta/2,30^{\circ})$ from $K=10^{4}$ heralded snapshots ($P_{d}=1$, photons shared equally), a trial counting as resolved when both estimates lie within $\Delta\theta/2$ of distinct true directions. SM-MUSIC resolves in $95$ of the $100$ trials at $\Delta\theta=20^{\circ}$ and in $100\%$ of the trials at $25^{\circ}$ and above, whereas FS-MUSIC reaches $0.24$ at $20^{\circ}$ and $0.55$ at $25^{\circ}$, its manifold bias pulling the peaks together, and needs $30^{\circ}$. MVDR-SM and Bartlett-SM are beamwidth-limited (HPBW $\approx30^{\circ}$ near $\theta=40^{\circ}$) and resolve only at $40^{\circ}$. No method resolves $\Delta\theta\leq12^{\circ}$: the second eigenvalue of the two-source $\rho_{R}$ is then $0.094$, against a noise-eigenvalue spread of about $\pm2\sqrt{N/K}=\pm0.08$. These values are specific to this array, criterion and snapshot budget.

\section{Robustness}\label{sec:robust}
Fig.~\ref{fig:robust} tests the receiver against the three imperfections that matter most for real dots. In every case the \emph{nominal} SM manifold, built for the ideal array, is used on purpose, so that the figure shows the effect of an unmodelled mismatch; errors are computed from the exact covariance at $P_{d}=0.1$. Pure dephasing of all 32 dots (as shown in Fig.~\ref{fig:robust}(a)) is the most instructive case. The SM model is given the broadened photon linewidth $\gamma+2\gamma_{\phi}$ but not the dephasing of the RX dots. At the resonant bound $\gamma_{\phi}=0.167\Gamma$ the SM error is $0.37^{\circ}$ against $1.33^{\circ}$ for FS, a factor $3.6$, with fidelity $0.774$ to the model. RX dephasing destroys the inter-dot coherence through which $\bm{G}_{RR}$ distorts the manifold (the purity of $\rho_{R}$ falls to $0.52$), so the received state drifts back toward the uncoupled form. The FS \emph{improves} while SM, which assumes coherent coupling, degrades, the curves crossing near $\gamma_{\phi}\approx0.8\Gamma$, and at the non-resonant $2.24\Gamma$ FS errs by $0.45^{\circ}$ against $1.04^{\circ}$. Coupling-aware processing thus needs resonantly excited, near-transform-limited dots; including RX dephasing in \eqref{eq:RSM} through the full Liouvillian of \eqref{eq:Cdot} is a direct extension.\\
The RX position disorder of Fig.~\ref{fig:robust}(b) is Gaussian with standard deviation $\sigma_{r}$ per coordinate, raises the SM error (RMS over 30 realizations) to $0.39^{\circ}$, $0.77^{\circ}$ and $1.97^{\circ}$ at $\sigma_{r}=0.01\lambda_{0}$ (\SI{15.5}{\nano\meter}), $0.02\lambda_{0}$ and $0.05\lambda_{0}$, while the FS stays between $1.68^{\circ}$ and $2.36^{\circ}$, so nanometer-scale knowledge of the element positions, by fabrication tolerance or by calibration, matters for keeping the uncalibrated error below one degree; the simulations sample discrete disorder levels and do not establish a sharp threshold.\\
The RX spectral disorder of  of Fig.~\ref{fig:robust}(c) denotes Gaussian detunings of standard deviation $\sigma_{\delta}$, is the most demanding imperfection and the main practical limitation: $\sigma_{\delta}=0.1\Gamma$ ($\approx\SI{13}{\mega\hertz}$) already raises the nominal-model SM error to $1.07^{\circ}$, at $\sigma_{\delta}=\Gamma$ SM and FS perform similarly ($6.3^{\circ}$ and $6.1^{\circ}$), and at $2\Gamma$ both fail, detunings of order $\Gamma$ reshaping the collective eigenmodes of $\Mm_{RR}$ much as a finite photon bandwidth does. Position disorder and spectral disorder are both static, however, so both can be calibrated: with the SM manifold built from the \emph{calibrated} $\Mm_{RR}$ and positions (10 realizations), the error returns to $0.032^{\circ}$ and $0.040^{\circ}$ for $\sigma_{r}=0.02\lambda_{0}$ and $0.05\lambda_{0}$, and to $0.048^{\circ}$ and $0.072^{\circ}$ for $\sigma_{\delta}=0.2\Gamma$ and $\Gamma$. In practice $\Mm_{RR}$ follows from single-dot spectroscopy with pairwise collective-decay measurements \cite{tiranov2023}; such calibration belongs to the receiver exactly as mutual-coupling calibration does for classical arrays \cite{friedlander1991}, and without it the dots must be tuned to a small fraction of $\Gamma$.

\section{Classical--Quantum Comparison and Limits}\label{sec:cq}
First-order radiation is classical-equivalent. Every pattern of Section~\ref{sec:results}-C is reproduced by a phased array with $w_{n}=c_{n}$, so state preparation replaces phase shifters without creating new patterns. \emph{Second-order statistics are genuinely non-classical}: $g^{(2)}\ll1$ in the beam (Table~\ref{tab:g2}) demonstrates strong antibunching and is consistent with single-photon emission under the source model used here. The DOA problem deals with finite bandwidth combined with RX coupling. The baseline shows a narrowband source and a single photon to be equally accurate on an uncoupled receiver, the difference on a coupled receiver being bandwidth alone (Fig.~\ref{fig:rxstate}). The SM correction is therefore not intrinsically quantum -- a classical receiver with the correct frequency-dependent coupled manifold can use it, and the improved two-source resolution (Section~\ref{sec:results}-H) results from removing manifold mismatch rather than from a new quantum-enhanced resolution mechanism. The one place where the single-photon character enters the bound is the measurement statistics: \eqref{eq:FIMexact} lies up to $27\%$ below the covariance-only Gaussian bound of a receiver with the same $\bm{R}_{x}$, an improvement that arises from the non-Gaussian statistics of the one-photon heterodyne model and that only a likelihood-based estimator can exploit. Both quantities are classical Fisher information bounds for this measurement, not quantum Fisher information.

\section{Discussion and Limitations}\label{sec:disc}
The bare link absorbs with probability $1.0\times10^{-4}$ at $R=200\lambda_{0}$, so the values of $P_{d}$ in Figs.~\ref{fig:rmse} and \ref{fig:resolution} describe retained gated or heralded snapshots rather than transmitted photons. In a fiber-coupled node \cite{yang2024,morrison2023}, $\bm{G}_{RT}$ would become the modal Green function of the fiber and couplers, while Proposition~\ref{prop:1} and the SM model, depending only on $\bm{h}(\omega)$ and $\Mm_{RR}$, carry over.\\
The dots are two-level systems with a single linear dipole, the orthogonal exciton lying outside the RX response band; phonon sidebands are lumped into $P_{d}$; the host is homogeneous, whereas a nanophotonic environment would enter through its own Green tensor in \eqref{eq:JG}; and the Markov elimination limits the model to $\Gamma R/c\ll1$ (Section~\ref{sec:channel}). The readout assumes one equal-efficiency collection channel per dot and traces out time, so a time-resolved heterodyne readout would expose the spectral structure of $\rho_{R}$ and invite a matched temporal-mode analysis. The nominal model assumes identical, frequency-tuned dots, which Section~\ref{sec:robust} shows to be essential unless calibrated. Two reductions should be read with their stated scope. The SM manifold uses the single-resonance Lorentzian approximation of $|\kappa(\omega)|^{2}$; Section~\ref{sec:results}-A quantifies that approximation and shows that it accounts for roughly half of the residual SM bias and none of the SM--CA gap, and the closed form can be replaced by quadrature of \eqref{eq:RSM} with the measured or computed spectrum where that residual matters. Receiver dephasing, in turn, enters here only through the photon linewidth in the coherent resolvent; a full treatment requires replacing the coherent resolvent of \eqref{eq:RSM} by a Liouvillian-based frequency-domain state calculation. All results reported in the present work are obtained numerically only, with future scope of extension for experimental validation. 

\section{Conclusion}\label{sec:concl}
A telecom C-band single-photon link between two $4\times4$ InAs/InP quantum-dot arrays has been modelled as one single photon-excitation Green's function problem with measured data of emitter parameters from the existing literature. The transmitter results provide a validation rather than a claim. The first-order steering array is equivalent to a classical phased array, with a pointing displacement up to $5.3^{\circ}$ and a $6.2$~dB directivity loss at the measured non-resonant dephasing, its antibunching being the principal nonclassical distinction in this model. On reception, the resolvent form with the full excitation vector is exact within the model, and frequency by frequency the single-photon response has the spatial shape of a coherent-drive response (Proposition~\ref{prop:1}); yet the finite bandwidth of a lifetime-limited photon leaves the coupled receiver, conditioned on absorption, in a mixed spatial state of purity $0.949$, and the narrowband coupling-aware manifold then errs by $1.81^{\circ}$, more than no correction. The spectrally matched manifold (SM), evaluated in closed form, lowers the model-mismatch error to $0.031^{\circ}$. In finite-snapshot simulations the SM estimators approach the corresponding CRBs, while the exact non-Gaussian CRB remains below the Gaussian covariance-based bound. The insight for array design is therefore that the finite bandwidth of a single photon, acting together with the frequency-selective collective coupling of the receiving array, leaves the receiver in a mixed spatial state, and it is that mixedness -- not the photon statistics -- that invalidates a narrowband manifold and motivates a spectrally matched one. Resonant excitation and in-situ calibration of the receiver coupling matrix are required; with both, realistic position and spectral disorder are tolerated. At equal numbers of detected photons a classical receiver supplied with the same frequency-dependent manifold would perform identically, and the single-photon regime is of interest for what it enables in a quantum network rather than for raw performance.

\bibliographystyle{IEEEtran}
\bibliography{refs}

\vspace{-2em}

\end{document}